\documentclass[a4paper,11pt]{article}
\pdfoutput=1 

\usepackage{jcappub} 

\usepackage[T1]{fontenc} 

\usepackage{graphicx,caption,subcaption,bm,color,hyperref}
\usepackage[normalem]{ulem} 
\usepackage{amsmath,amssymb}
\usepackage{mathtools} 
\usepackage{dsfont} 
\usepackage{multirow} 
\usepackage{float} 
\usepackage{mathrsfs} 
\usepackage{tensor} 
\usepackage{booktabs}
\usepackage{ragged2e}
\usepackage{tabularx} 
\usepackage[british]{babel}
\usepackage{orcidlink}
\usepackage[dvipsnames]{xcolor}
\usepackage{arydshln}

\hypersetup{colorlinks, linkcolor={blue}, citecolor={red}, urlcolor={blue}}

\definecolor{amaranth}{rgb}{0.9, 0.17, 0.31}
\definecolor{palatinateblue}{rgb}{0.15, 0.23, 0.89}
\definecolor{brightpink}{rgb}{1.0, 0.0, 0.5}
\definecolor{brightgreen}{rgb}{0.14, 0.84, 0.72}
\definecolor{mediumgreen}{rgb}{0.22, 0.67, 0.59}
\definecolor{darkgreen}{rgb}{0.25, 0.5, 0.46}
\definecolor{verydarkgreen}{rgb}{0.22, 0.33, 0.31}

\definecolor{myRed}{HTML}{E64248}
\definecolor{myPink}{HTML}{C57B9A}
\definecolor{myPurple}{HTML}{4C72B0}

\definecolor{posurl}{rgb}{0.1, 0.3, 0.75}
\definecolor{posurl_blue}{rgb}{0.1, 0.3, 0.9}
\definecolor{posurl_dark}{rgb}{0.1, 0.4, 0.75}

\hypersetup{ linktoc=all,
    colorlinks, linkcolor={posurl},
    citecolor={posurl_dark}, urlcolor={posurl_blue}
}
\DeclareUnicodeCharacter{03A9}{$\Omega$}
\DeclareUnicodeCharacter{039B}{$\Lambda$}

\graphicspath{{images/}}

\newcommand{\be}{\begin{equation}}
\newcommand{\ee}{\end{equation}}
\newcommand{\ba}{\begin{eqnarray}}
\newcommand{\ea}{\end{eqnarray}}

\newcommand{\erf}{\operatorname{erf}}

\begin{document}

\title{\boldmath Alleviating the Hubble Tension with Smooth Sign-Switching Dark Energy: Full CMB Constraints with DESI and PantheonPlus}

\author[a,b,c]{Mariam Bouhmadi-López 
\orcidlink{https://orcid.org/0000-0002-1529-1889} }
\emailAdd{mariam.bouhmadi@ehu.eus}

\author[d]{,Hsu-Wen Chiang 
\orcidlink{https://orcid.org/0000-0002-5450-9297} }
\emailAdd{jiangxw@sustech.edu.cn, b98202036@ntu.edu.tw}

\author[b]{and Beñat Ibarra-Uriondo\,
\orcidlink{https://orcid.org/0009-0008-8064-2340} }
\emailAdd{benat.ibarra@ehu.eus}

\affiliation[a]{IKERBASQUE, Basque Foundation for Science, 48011, Bilbao, Spain}
\affiliation[b]{Department of Physics, University of the Basque Country UPV/EHU, P.O. Box 644, 48080 Bilbao, Spain}
\affiliation[c]{EHU Quantum Center, University of the Basque Country UPV/EHU, P.O. Box 644, 48080 Bilbao, Spain}
\affiliation[d]{Department of Physics, Southern University of Science and Technology, Shenzhen 518055, China}

\abstract{

Sign-switching dark energy has recently been proposed as a minimal modification of the late-time expansion history aimed at alleviating tensions within the standard cosmological model. In this work, we investigate ECDM, a smooth realisation of this scenario, with the dark energy density gradually transitioning from a negative to a positive value. We develop a consistent formulation of the perturbation equations that remains well behaved even when the dark energy equation-of-state parameter diverges during the transition. We confront the model with a comprehensive set of cosmological observations, including cosmic microwave background measurements from Planck 2018, ACT DR6 and SPT-3G, baryon acoustic oscillation measurements from DESI DR2, Type Ia supernova distances from Pantheon+, and local measurement of the Hubble constant from SH0ES. The inclusion of perturbations allows us to assess the impact of the model on structure growth and CMB anisotropies, providing a more thorough test of sign-switching dark energy. Our results show that this class of models is fully compatible with current precision cosmological observations while alleviating the Hubble tension and providing a compelling modification of the late-time dynamics of the Universe.

}

\maketitle

\tableofcontents

\section{Introduction}\label{intro}

In recent decades, a wealth of cosmological observations has established that the expansion of the Universe is currently undergoing accelerated growth, driven by a component with sufficiently negative pressure that becomes dominant at late times. The first direct evidence for this accelerated expansion came from observations of Type Ia supernovae (SN{e}~Ia)~\cite{SupernovaSearchTeam:1998fmf,SupernovaCosmologyProject:1998vns}, whose luminosity distance measurements revealed a transition in the expansion history from decelerating to accelerating. This picture is further supported by observations of the cosmic microwave background (CMB)~\cite{Planck:2018vyg,AtacamaCosmologyTelescope:2025blo,SPT-3G:2025bzu,SPT-3G:2025vyw} and baryon acoustic oscillations (BAO) in the large-scale distribution of galaxies~\cite{Eisenstein2005,Cole2005,Percival2010,Tegmark2004,Reid2010,eBOSS:2020yzd,Scolnic:2021amr,Brout:2022vxf,Rubin:2023jdq,DESI:2022lza,DESI:2024aqx,DESI:2024jxi,DESI:2024lms,DESI:2024lzq,DESI:2024mwx,DESI:2024uvr,Elbers:2025vlz,DESI:2025fii,DESI:2025zgx,Sabogal:2025qhz}. Additional evidence is provided by cosmic chronometer (CC) measurements~\cite{Jimenez:2001gg,Moresco:2024wmr} and local determinations of the Hubble constant, $H_0$~\cite{H0DN:2025lyy,Freedman:2024eph}. 

These observations are remarkably well described within the framework of General Relativity (GR), whose simplest cosmological realisation is the standard $\Lambda$CDM model. In this framework, the late-time accelerated expansion is driven by a cosmological constant, $\Lambda$, which acts as dark energy (DE) and is characterised by an equation-of-state (EoS) parameter $w=-1$. Despite its observational success, the $\Lambda$CDM model faces several theoretical and observational challenges. On the theoretical side, the cosmological constant suffers from the well-known fine-tuning problem, arising from the enormous discrepancy between the observed value of the vacuum energy density and theoretical expectations from quantum field theory. Closely related is the cosmic coincidence problem, which questions why the energy densities of matter and DE are of the same order of magnitude precisely at the present epoch, despite their vastly different evolutionary histories. These issues suggest that $\Lambda$CDM should perhaps be regarded as an effective phenomenological description rather than the fundamental theory. From an observational perspective, increasing attention has been drawn to tensions between early- and late-Universe probes. The most prominent example is the Hubble tension, namely the discrepancy between the value of $H_0$ inferred from CMB observations assuming $\Lambda$CDM and that obtained from local distance ladder measurements, which currently exceeds the $5\sigma$ level. Such discrepancies have motivated extensive investigations of extensions to the standard cosmological model, particularly those involving modifications to the dark energy sector, with the aim of constructing theoretically motivated scenarios capable of alleviating observational tensions while remaining consistent with the full range of cosmological data~\cite{Weinberg:1988cp,Sahni:1999gb,Sahni:2002kh}.

The theoretical challenges faced by the $\Lambda$CDM model, together with the emergence of observational tensions, have motivated the development of numerous extensions to the standard cosmological paradigm \cite{CosmoVerseNetwork:2025alb}. One of the most widely explored avenues consists of replacing the cosmological constant with a dynamical dark energy (DDE) component whose properties evolve throughout cosmic history. In this context, scalar fields provide a natural framework for constructing time-dependent DE models, 
ranging from canonical quintessence scenarios \cite{PhysRevD.37.3406} to non-canonical realisations such as $k$-essence \cite{PhysRevD.63.103510}. Axion-inspired models have also attracted considerable attention owing to their strong theoretical motivation from high-energy physics and string-theoretic constructions \cite{Kamionkowski:2014zda,Emami:2016mrt,Chiang:2025qxg}. Beyond scalar fields, effective fluid descriptions \cite{Kamenshchik:2001cp}, $p$-form fields \cite{Koivisto:2009fb,Koivisto:2009ew,Morais:2016bev,Bouhmadi-Lopez:2016dzw,Bouhmadi-Lopez:2025lzm,Bouhmadi-Lopez:2026ytj,Heisenberg:2014rta,deRham:2020yet,DeFelice:2020sdq,Chiang:2025hrj}, and interacting dark sector scenarios \cite{Li:2025owk,Li:2026xaz} have been investigated as viable mechanisms for generating the observed late-time acceleration. An alternative possibility is that cosmic acceleration does not arise from an additional DE component, but instead reflects a departure from GR on cosmological scales. This idea has led to a broad spectrum of modified gravity theories that alter the gravitational dynamics while reproducing the successful predictions of GR in appropriate limits. Representative examples include bigravity theories \cite{PhysRevD.3.867,DeFelice:2020ecp}, scalar--tensor models with derivative self-interactions such as kinetic gravity braiding \cite{Deffayet:2010qz,Pujolas:2011he,BorislavovVasilev:2022gpp,BorislavovVasilev:2024loq}, and geometric extensions based on curvature, torsion, or non-metricity. Well-known realisations of these ideas include $f(\mathcal{R})$ \cite{Sotiriou:2008rp,Capozziello:2011et,Nojiri:2010wj,Bouhmadi-Lopez:2010qyi,Nojiri:2017ncd,Morais:2015ooa}, $f(\mathcal{T})$ \cite{Bengochea:2008gz,Ferraro:2006jd,Cai:2015emx,Bouhmadi-Lopez:2026dte}, and $f(\mathcal{Q})$ gravity \cite{BeltranJimenez:2018vdo,BeltranJimenez:2019tme,Hashim:2020sez,Hashim:2021pkq,Ayuso:2020dcu,Boiza:2025xpn,Ayuso:2025vkc}, together with more general frameworks incorporating boundary terms that connect different geometric formulations of gravity, such as $F(Q,\mathcal{B})$ \cite{Capozziello:2023vne,De:2023xua}, $f(T,\mathcal{B})$ \cite{Bahamonde:2015zma,Bahamonde:2019shr}, and $f(\mathcal{R},T)$ theories \cite{Lamaaoune:2025zhd}. 

Among the broad landscape of DDE models, increasing attention has recently been directed towards scenarios in which the DE density changes sign during the late-time evolution of the Universe~\cite{Akarsu:2021fol,Akarsu:2022typ,Akarsu:2023mfb,Paraskevas:2024ytz,Yadav:2024duq,Akarsu:2024qsi,Akarsu:2024eoo,Akarsu:2024nas,Souza:2024qwd,Akarsu:2025gwi,Akarsu:2025dmj,Akarsu:2025ijk,Escamilla:2025imi,Akarsu:2025nns,Anchordoqui:2023woo,Anchordoqui:2024gfa,Anchordoqui:2024dqc,Soriano:2025gxd}. In these models, the effective DE sector evolves from a negative-energy phase, corresponding to an anti-de Sitter (AdS)-like plateau, to a positive-energy de Sitter (dS) state at low redshifts. Such a transition modifies the recent expansion history while leaving the sound horizon at the drag epoch, $r_{\rm d}$, essentially unaffected, provided that the pre-recombination physics remains unchanged. This characteristic makes sign-switching DE models particularly appealing, as they can alter late-time cosmological observables without modifying the physics of the early Universe. The first indication of this behaviour emerged in the context of the graduated dark energy model~\cite{Akarsu:2019hmw}, and was subsequently extended to the $\Lambda_{\rm s}$CDM framework and related cosmological scenarios. More generally, analogous phenomenology has been realised in a variety of frameworks, including omnipotent dark energy models and other constructions featuring sign-changing DE densities~\cite{DiValentino:2020naf,Adil:2023exv,Specogna:2025guo}. Owing to their ability to modify the late-time expansion history while preserving the standard early-Universe picture, these models have been extensively investigated as possible extensions of the concordance cosmological model. The broader possibility that the DE density may become negative during part of cosmic history has been explored from a variety of theoretical and phenomenological perspectives, including vacuum-energy extensions, string landscape, interacting dark sector scenarios, effective fluid descriptions, and other DDE models~\cite{AlbertoVazquez:2012ofj,Sahni:2002dx,Vazquez:2012ag,BOSS:2014hwf,Sahni:2014ooa,BOSS:2014hhw,DiValentino:2017rcr,Mortsell:2018mfj,Poulin:2018zxs,Wang:2018fng,Banihashemi:2018oxo,Dutta:2018vmq,Banihashemi:2018has,Akarsu:2019ygx,Li:2019yem,Visinelli:2019qqu,Perez:2020cwa,Akarsu:2020yqa,Ruchika:2020avj,Yang:2020ope,Calderon:2020hoc,DeFelice:2020cpt,Paliathanasis:2020sfe,Bonilla:2020wbn,Acquaviva:2021jov,Bag:2021cqm,Bernardo:2021cxi,Escamilla:2021uoj,Sen:2021wld,Ozulker:2022slu,DiGennaro:2022ykp,Akarsu:2022lhx,Moshafi:2022mva,vandeVenn:2022gvl,Ong:2022wrs,Tiwari:2023jle,Malekjani:2023ple,Ben-Dayan:2023rgt,Vazquez:2023kyx,Alexandre:2023nmh,Adil:2023ara,Paraskevas:2023itu,Gomez-Valent:2023uof,Wen:2023wes,DeFelice:2023bwq,Menci:2024rbq,Gomez-Valent:2024tdb,DESI:2024aqx,Bousis:2024rnb,Wang:2024hwd,Colgain:2024ksa,Tyagi:2024cqp,Toda:2024ncp,Sabogal:2024qxs,Dwivedi:2024okk,Escamilla:2024ahl,Pai:2024ydi,Wen:2024orc,Gomez-Valent:2024ejh,Manoharan:2024thb,Mukherjee:2025myk,Efstratiou:2025xou,Tamayo:2025xci,Wang:2025dtk,Gonzalez-Fuentes:2025lei,Bouhmadi-Lopez:2025spo,Bouhmadi-Lopez:2025ggl,Hogas:2025ahb,Gomez-Valent:2025mfl,Tan:2025xas,Yadav:2025vpx,Pedrotti:2025ccw,Forconi:2025gwo,Nyergesy:2025lyi,Ghafari:2025eql,Akarsu:2026anp,Gokcen:2026pkq,Banks:2003es,2026ForPh..7470073L}. Taken together, these studies demonstrate that current cosmological observations do not require the DE density to remain strictly positive at all epochs. Instead, they permit richer evolutionary histories than those predicted by a cosmological constant. In particular, model-independent reconstructions of interacting dark energy kernels have found no compelling evidence against negative effective DE densities at redshifts $z \gtrsim 2$~\cite{Escamilla:2023shf}.

In this work, we move beyond the abrupt transition approximation of the $\Lambda_s$CDM framework and consider an extension in which the transition occurs smoothly, with controllable transition speed. To this end, we adopt the ECDM model introduced in \cite{Bouhmadi-Lopez:2025ggl,Bouhmadi-Lopez:2025spo,Bouhmadi-Lopez:2026ckz}, in which the DE density is described by an error-function profile that allows for a continuous interpolation between an early-time negative density and a late-time positive de Sitter-like phase. While our previous analyses of this framework have focused primarily on background-level observable constraints \cite{Ibarra-Uriondo:2026zbp}, in this work, we extend the study by incorporating consistent perfect-fluid-based linear perturbations and confronting the model with a comprehensive set of observational data. In particular, we perform a joint analysis using CMB measurements from Planck 2018 low-$\ell$ temperature anisotropies ~\cite{Planck:2018vyg}, ACT DR6 CMB-only likelihoods (including both ACT and ACT–Planck high-$\ell$ cross-spectra)~\cite{AtacamaCosmologyTelescope:2025blo}, and SPT-3G data~\cite{SPT-3G:2025bzu}, together with late-time geometric distance probes from the DESI DR2 BAO~\cite{DESI:2025zgx} and the PantheonPlus+SH0ES SNe~Ia~\cite{Brownsberger:2021uue}. Additionally, we include CMB lensing information from the ACT DR6 lensing likelihood, thereby improving constraints on the growth of structure, as well as high-resolution polarisation measurements from SPT-3G two-year data. 

A central technical difficulty in this class of models arises from the fact that the effective DE EoS becomes ill-defined across the sign transition of the energy density, which complicates the standard treatment of perturbations. To overcome this issue, we reformulate the perturbation equations in a manifestly well-defined form that remains regular throughout the entire cosmological evolution, allowing for a stable numerical implementation within the Boltzmann solver framework. This enables us to consistently propagate perturbations through the transition regime and to confront the model with full CMB power spectra and lensing data. Furthermore, we exploit this framework to assess the sharpness and effective duration of the DE transition, examining whether the smooth error-function behaviour preferred by the model corresponds to a rapid or extended interpolation between the AdS-like and dS-like regimes.

In addition to the observational fit of cosmological observables, we further extend our study to the analysis of the CMB temperature, polarisation, and lensing power spectra.
In particular, we compute the linear matter power spectrum in order to assess the impact of the sign-switching DE dynamics on the clustering of matter at large scales. We also examine the full CMB angular power spectra, including their sensitivity to late-time integrated effects, with particular emphasis on the integrated Sachs--Wolfe (ISW) contribution, which provides a direct probe of time-varying gravitational potentials and is therefore especially sensitive to deviations from $\Lambda$CDM at low redshifts \cite{Crittenden:1995ak,Zhao:2008bn,dePutter:2010vy}. Moreover, we analyse the weak lensing potential and its corresponding angular power spectrum, which encode complementary information on the matter distribution integrated along the line of sight and offer an additional consistency check on the growth history inferred from primary CMB anisotropies \cite{Giannantonio:2008zi,Stolzner:2017ged,Kovacs:2021mnf}. Taken together, these observables provide a comprehensive framework to test the viability of the model across a wide range of cosmological scales and physical processes.

The structure of this paper is as follows. In section~\ref{sec:cosmology} we review the late-time background dynamics of the Universe. Section~\ref{sec:model} introduces the ECDM model and its smooth transition between negative and positive DE densities. Section~\ref{sec:metric} presents the formulation of linear cosmological perturbations and the rescaled density contrast for the sign-switching DE. Section~\ref{sec:methodolofy} is devoted to describing the datasets and the methodology employed in the MCMC analysis. Section~\ref{sec:observations} presents the main results of our study, including constraints on cosmological parameters, statistical analysis, and the reconstructed expansion history. Section~\ref{sec:powerspectra} extends the analysis to derived observables, including the matter power spectrum, CMB temperature and polarisation spectra, and CMB lensing and ISW effect. Finally, Section~\ref{sec:conclusions} summarises our conclusions and discusses future prospects for sign-switching DE models. Technical details are provided in the Appendices: Appendix~\ref{appendix:CPT} and \ref{appendix:synchronous} review the cosmological perturbation theory and present the modified perturbation equations in the synchronous gauge, and Appendix~\ref{appendix:statistics} the statistical definitions used in this work.

The simulations performed in this work follow a methodology closely aligned with the specifications of the COST Action CA21136 “Addressing observational tensions in cosmology with systematics and fundamental physics” (CosmoVerse) Cosmology Compilation Group, to be released later this year. The main difference lies in the implementation strategy: whereas the CosmoVerse pipeline employs parallelisation to accelerate the likelihood evaluations, we instead perform the analyses using independent Markov chains.

\section{Late-time Cosmology\label{sec:cosmology}}

In this section, we provide a general overview of the late-time evolution of the Universe. Let us consider a homogeneous and isotropic Universe described by the Friedmann-Lema\^{i}tre-Robertson-Walker (FLRW) metric. For a spatially flat Universe ($\mathcal{K}=0$), the Friedmann and Raychaudhuri equations are:
\begin{equation}
H^2 = \frac{\kappa^2}{3} \sum_A \rho_A  \,,\quad
\frac{\ddot a}{a} = -\frac{\kappa^2}{6} \sum_A \left( \rho_A + 3p_A \right)  \,,
\end{equation}
where $H \equiv \dot{a}/a$ is the Hubble parameter, a dot denotes differentiation with respect to cosmic time, the subscript $A$ enumerates over species of fluids, and $\kappa^2 \equiv 8\pi G$ with $G$ the gravitational constant. The total energy density and pressure are denoted by $\rho$ and $p$, respectively.

Following a multi-fluid approach we consider a universe composed by radiation, non-relativistic matter (cold dark matter and baryons) and DE, with
\begin{equation}
\rho = \rho_{\rm r} + \rho_{\rm m} + \rho_{\rm d}  \,,\quad
p = p_{\rm r} + p_{\rm m} + p_{\rm d}  \,,
\end{equation}
where the subscripts $\rm r$, $\rm m$ and $\rm d$ denote radiation, matter and DE, respectively. We assume no direct interaction between components, so each separately satisfies the conservation equation:
\begin{equation}
\dot{\rho}_A + 3H (\rho_A + p_A) = 0  \,.
\end{equation}

The equation of state parameters are defined as:
\begin{equation}
w_{\rm r} \equiv \frac{p_{\rm r}}{\rho_{\rm r}} = \frac{1}{3}  \,,\quad
w_{\rm m} \equiv \frac{p_{\rm m}}{\rho_{\rm m}} = 0  \,,\quad
w_{\rm d} \equiv \frac{p_{\rm d}}{\rho_{\rm d}}  \,,
\end{equation}
where the DE EoS parameter is allowed to evolve with redshift. From the conservation equation, the radiation and matter density scales respectively as $\rho_{\rm r} = \rho_{\rm r0} (1+z)^4$ and $\rho_{\rm m} = \rho_{\rm m0} (1+z)^3$, while the DE density evolves as:
\begin{equation}
\rho_{\rm d}(z) = \rho_{{\rm d}0} \, \exp\left( 3 \int_0^z \frac{1 + w_{\rm d}(z')}{1+z'} dz' \right)  \,,
\end{equation}
where a $0$-subscript denotes the present-day value of a quantity. The Friedmann equation then becomes:
\begin{equation}
\frac{H^2}{H_0^2} = \Omega_{\rm r0} (1+z)^4 + \Omega_{\rm m0} (1+z)^3 + \Omega_{{\rm d}0} \, \frac{\rho_{\rm d}}{\rho_{{\rm d}0}}  \,,
\label{eq:H_late}
\end{equation}
with $H_0$ the Hubble constant, $\Omega_A = \rho_A/\rho_{\rm c}$ the fractional density, $\rho_{\rm c} = 3H^2/\kappa^2$ the critical density. The present-day fractional densities satisfy $\Omega_{\rm r0}+\Omega_{\rm m0} + \Omega_{{\rm d}0} = 1$ for a flat Universe. The total EoS parameter of the cosmic fluid can be expressed as:
\begin{equation}
w_{\rm tot}(z) \equiv \frac{p}{\rho} = \Omega_{\rm r} w_{\rm r} + \Omega_{\rm m} w_{\rm m} + \Omega_{\rm d} w_{\rm d}  \,.
\label{eq:wtot}
\end{equation}

\section{Model description\label{sec:model}}

The model considered in this work, which belongs to a subclass of the $\Lambda_{\rm s}$CDM framework and is commonly referred to as ECDM  \cite{Bouhmadi-Lopez:2025ggl,Bouhmadi-Lopez:2025spo,Bouhmadi-Lopez:2026ckz}, is constructed by means of an interpolating error-function profile.
More specifically, the DE density is parametrised in $x \equiv \log a$ as
\begin{equation}
\rho_{\rm d}(x) = \rho_{\rm d0} \frac{ \erf \!\left( \eta ( x - x_\dagger ) \right) }{ \erf ( - \eta x_\dagger ) }  \,,
\label{densityecdm}
\end{equation}
where $\erf(x) = \frac{2}{\sqrt\pi} \int_0^x e^{-y^2} dy$ is the error function, and $\eta$ controls the rapidity of the transition. In fact, the larger is $\eta$ the sharper is the transition, while the smaller is $\eta$ the smoother is the transition, while $x_\dagger = -\ln(1 + z_\dagger)$ sets the characteristic redshift at which the sign change occurs. Here, $z_\dagger$ denotes the redshift of the transition and represents an additional free parameter beyond those of the standard $\Lambda$CDM model. Consequently, the ECDM scenario constitutes a two-parameter extension of $\Lambda$CDM, with $\{ \eta, z_\dagger \}$ governing the dynamics of the DE sector. For completeness, we recall that the original $\Lambda_{\rm s}$CDM; i.e. a cosmological model with a sharp transition between a positive and a negative cosmological constant of equal magnitude at the late-time redshift $z_\dagger$, when $\eta\rightarrow\infty$, we recover such transition. In addition, $\Lambda$CDM is recovered when $z_\dagger\rightarrow\infty$.

Regarding the EoS parameter, it can be readily obtained by imposing the conservation of the energy momentum tensor of DE:
\begin{equation}
w_\mathrm{d} = -1 - \frac{ 2 \eta e^{-\eta^2 \left( x-x_\dagger \right)^2} }{3\sqrt\pi \erf \!\left( \eta (x - x_\dagger) \right)}  \,.
\label{eosecdm}
\end{equation}

Due to the continuous transition exhibited by the DE density at the sign-switching point $x = x_\dagger$, the EoS parameter diverges, undergoing a crossing of the cosmological constant line through infinite values. Nonetheless, the total EoS parameter remains well defined as
\begin{equation}
w = \frac{1}{3}\Omega_{\rm r} - \frac{\Omega_{\rm d0}}{E^2} \left( 1 + \frac{2\eta e^{-\eta^2 \left( x-x_\dagger \right)^2}}{3\sqrt\pi \erf(-\eta x_{\dagger})} \right)  \,,
\end{equation}
where $E=H/H_0$ is the normalised Hubble parameter. A key subtlety of this construction is that the DE EoS parameter does not, in general, constitute a globally well-defined or reliable diagnostic of the dynamics~\cite{Akarsu:2025gwi,Akarsu:2026anp,Gokcen:2026pkq}. In particular, in scenarios in which the DE density crosses the $\rho_{\rm d}=0$ boundary, as is the case in sign-switching DE models, the EoS parameter becomes ill-defined, despite the fact that the background cosmological evolution remains smooth and regular throughout the transition.

\section{Cosmological perturbation in Newtonian gauge\label{sec:metric}}

We have thus far described a perfectly homogeneous and isotropic universe. This constitutes an idealisation; in practice, our interest lies in the departures from exact homogeneity and isotropy that enable us to characterise and distinguish between cosmological models. To address this, we consider small perturbations about the background spacetime, assuming that these perturbations form a statistically homogeneous and isotropic distribution.

\subsection{Perturbation equation\label{subsec:newtonian}}

In the Newtonian gauge\footnote{For details of the cosmological perturbation theory, c.f. Appendix~\ref{appendix:CPT}. Here we select Newtonian gauge for simplicity. However, the common Einstein--Boltzmann codes \texttt{CAMB} and \texttt{CLASS} are formulated in the synchronous gauge, as detailed in Appendix~\ref{appendix:synchronous}.}, the perturbed FLRW metric is expressed in terms of the Bardeen potentials $\Phi$ and $\Psi$\cite{Bardeen:1980kt}:
\begin{equation}
d s^2=a^2\left[-\left(1+2 \Phi\right) d \tau^2+\left(1-2 \Psi\right)\delta_{ij} d x^i dx^j\right]  \,,
\label{gauge1}
\end{equation}
where $\tau \equiv \int a^{-1}dt$ is the conformal time and the Latin indices enumerate over Cartesian coordinates of the 3-space.

In the absence of anisotropic stress, the Bardeen potentials coincide, i.e. $\Psi=\Phi$. By approximating the constituent of the universe as multiple non-interacting fluids, the evolution of the fractional energy density perturbations $\delta_A\equiv \delta \rho_{A} / \rho_{A}$ and the velocity $v_A$, with $( \rho_A + P_A ) \partial^i v_A \equiv \delta\left(T_A\right)^i_{\,0}$ the perturbed momentum, is governed by the following set of first-order differential equations in Fourier space for each component fluid $A$, derived from the conservation of the perturbed energy-momentum tensor:
\be\begin{aligned}
\delta_A' &=  3 \mathcal{H} \left(w_A-c_{s A}^2\right) \delta_A + 3 \left(1+w_A\right) \Psi'  + \left(1+w_A\right) \left[ 9 \mathcal{H}^2 \left( c_{s A}^2 - c_{a A}^2 \right) - k^2 \right] v_A  \,,\\
v_A' &= \left(3 c_{s A}^2 - 1\right) \mathcal{H} v_A - \frac{c_{s A}^2}{1+w_A} \delta_A - \Psi  \,,
\label{eq:pertubations_general}
\end{aligned}\ee
where the superscript prime denotes the differentiation with respect to the conformal time, $\mathcal H$$\equiv a' / a = a H$ is the conformal Hubble parameter, $k$ is the wavelength of the Fourier modes, $c_{sA}^2 \equiv \delta p_A/\delta\rho_A$ is the rest-frame speed of sound, and $c_A^2 \equiv p'_A/\rho'_A = w_A - w_A' / (3\mathcal H ( 1 + w_A ))$ is the adiabatic speed of sound.

For components well approximated by the barotropic fluid such as the dark matter and the radiation in the late universe, both speeds of sound approach the EoS parameter, but not necessarily so for DE. In ECDM, we have
\begin{equation}
c_{a\rm d}^2 = \frac{2\eta^2 (x-x_\dagger)}{3} - 1  \,,
\end{equation}
which transits from a negative value to a positive value at $z = z_\dagger$. On the other hand, $c_{s{\rm d}}^2$ depends on the underlying model. We fix it to unity for simplicity.\footnote{We have verified that $c_{sA}^2$ has no adverse effect on both CMB power spectrum and matter power spectrum up to a small positive value.}

Nonetheless, this set of equations is not well defined for sign-switching DE models. In such scenarios, the EoS parameter diverges as DE density crosses zero, leading to a corresponding divergence in the fractional DE density perturbations $\delta_{\rm d}$. To circumvent this issue, one can instead consider the ratio between $\delta\rho_{\rm d}$ and $\rho_{\rm d} + P_{\rm d}$:
\begin{equation}
f_A \equiv \frac{\delta_A}{1 + w_A}  \,,
\label{eq:new_pert}
\end{equation}
which remains well defined throughout the transition and provides a consistent description of the perturbative dynamics.
Substituting $\delta_A$ with $f_A$ in Eq.~\eqref{eq:pertubations_general} we arrive at
\be\begin{aligned}
f_A' &= 3\mathcal{H} ( c_{a A}^2 -  c_{sA}^2 ) f_A +  3 \Psi'  + \left[ 9 \mathcal{H}^2 \left( c_{s A}^2 - c_{a A}^2 \right) - k^2 \right] v_A  \,,\\
v_A' &= ( 3 c_{s A}^2 - 1 ) \mathcal{H} v_A - c_{s A}^2 f_A - \Psi  \,.
\label{eq:newtonian_solved}
\end{aligned}\ee

In this new formulation we can compute the rescaled DE perturbation $f_A$ without encountering the pole at $\rho_A=0$. Notice that for the new quantity $f_A={\delta\rho_A}/{(\rho_A+p_A})$  the relevant normalisation is not the energy density itself but the inertial mass density, $\varrho_A = \rho_A + p_A$. Consequently, one may expect that a more thorough analysis of this quantity is required in the context of sign-switching scenarios.

\begin{figure}[htbp]
    \centering
    \includegraphics[width=\linewidth]{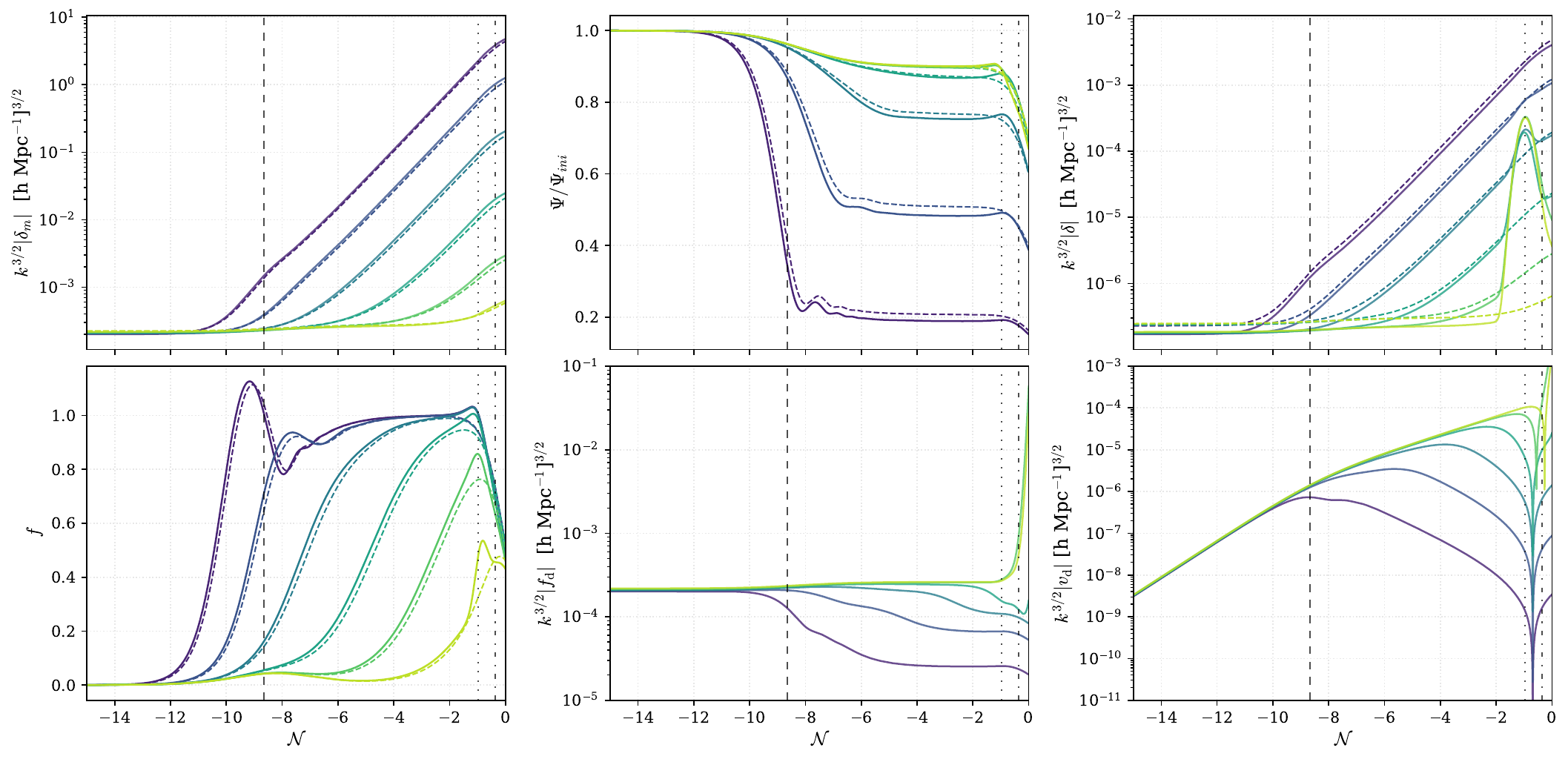}
    \caption{\justifying{
Comparison between $\Lambda$CDM (dashed lines) and the ECDM model (solid lines) for four linear perturbation quantities as a function of the e-folding time $\mathcal{N}=\ln a$ for the MAP values form the CMB-SPA combination (In the case of the ECDM model we choose the fast-transition branch). The panels show: (upper left) the matter density perturbations $k^{3/2}|\delta_m|$, (upper centre) the gravitational potential $\Psi/\Psi_{\mathrm{ini}}$, (upper right) the total density perturbations $k^{3/2}|\delta|\equiv k^{3/2} |\rho_{\rm c}^{-1} \sum\delta \rho_A|$, (lower left) the linear growth rate $f = \mathrm{d}\ln \delta_m / \mathrm{d}\mathcal{N}$, (lower centre) the DE perturbations $k^{3/2}|f_{\rm d}|$, where $f_{\rm d} \equiv \left( 1 + w_{\rm d} \right)^{-1} \delta_{\rm d}$ and (lower right) the DE velocity $k^{3/2}|v_d|$.  Each colour corresponds to a different Fourier mode:k = 3.33 $\times
10^{-4}$ h Mpc$^{-1}$ ({\color{SpringGreen}Yellow}), k = 1.04 $\times10^{-3}$ h Mpc$^{-1}$ ({\color{Green}Green-yellow}), k = 3.27$\times10^{-3}$ h Mpc$^{-1}$ ({\color{JungleGreen}Green}), k = $1.02 \times10^{-2}$ h Mpc$^{-1}$ ({\color{MidnightBlue}Cyan}), k = 3.19$\times10^{-2}$ h Mpc$^{-1}$ ({\color{BlueViolet}Blue}), k = 0.1 h Mpc$^{-1}$ ({\color{Fuchsia}Purple}). Vertical black lines indicate characteristic background transitions: radiation–matter equality (dashed), ECDM sign-switching point $x_\dagger$ (dotted), and matter--DE equality (dot-dashed). 
}}
    \label{fig:newtonia_gauge_plots6}
\end{figure}

\subsection{Perturbation evolution history}\label{subsec:pert_evolution}

Using the best fit parameters in Table~\ref{tab:new} we evolve the perturbations in Newtonian gauge for 15 e-folds. The results are shown in figure~\ref{fig:newtonia_gauge_plots6}. 

In the top-left image of figure~\ref{fig:newtonia_gauge_plots6} we have the evolution of the matter perturbations. We observe small deviations from the $\Lambda$CDM evolution across all Fourier modes, with the most pronounced differences occurring for modes that enter the horizon around the epoch of matter domination. Although the overall qualitative behaviour remains unchanged, these intermediate-scale modes display mild but systematic departures relative to the standard scenario. As in the standard picture, during radiation domination the matter density contrast $\delta_{\rm m}$ remains approximately constant for each mode until horizon crossing. After horizon entry, gravitational instability drives the growth of perturbations, which becomes increasingly significant throughout the matter-dominated era. This growth is subsequently suppressed once DE becomes dynamically important, leading $\delta_{\rm m}$ to approach an approximately constant late-time value for each mode. Overall, modes that enter the horizon earlier (larger $k$) follow a similar evolutionary pattern but exhibit slightly different amplitudes compared to $\Lambda$CDM, with the most visible enhancement concentrated in the range of modes that enter the horizon during matter domination.

In the top-centre image of figure~\ref{fig:newtonia_gauge_plots6}, we show the evolution of the gravitational potential, normalised to its initial value, $\Psi/\Psi_{\rm ini}$, for ECDM in comparison with $\Lambda$CDM, considering the same set of Fourier modes as in the previous case. In contrast to the results presented in \cite{Bouhmadi-Lopez:2025spo}, where a more abrupt suppression of the gravitational potential was found around the sign-switching epoch (particularly for small $k$ modes), the inclusion of DE perturbations in the present analysis leads to a smoother and less sharp transition. Overall, the smallest scales no longer exhibit the pronounced, sudden decay previously observed at the sign-change, although residual departures from $\Lambda$CDM remain visible. By contrast, larger $k$ modes display systematic suppression during radiation and matter domination compared to $\Lambda$CDM. As the DE component becomes dynamically relevant, $\rho_{\rm d} +P_{\rm d}$ boost the potential, specially for modes that enter the horizon around $z_\dagger$. 
Despite these differences, the late-time evolution converges to a behaviour consistent with that found in the earlier analysis, albeit with a smoother transition profile.

Regarding the growth rate, $f = \mathrm{d}\ln\delta_{\rm m}/\mathrm{d}\mathcal{N}$ observed in the bottom-left image of figure~\ref{fig:newtonia_gauge_plots6}, in contrast to the behaviour reported in \cite{Bouhmadi-Lopez:2025spo}, where a sharper feature was observed for small-scale modes at the sign change, the inclusion of perturbations leads to a smoother and more gradual evolution, with a more pronounced feature at large scale instead. Nevertheless, all modes exhibit the largest departure from the $\Lambda$CDM growth history at the transition, with a consistent, albeit milder, deviation from the mode re-entry till late time as DE comes to dominate.

A comment on the evolution of the DE perturbations is much required, as they have not been previously explored in this context. In the present model, the DE component remains nearly constant over most of the cosmic history (region when $w_{\rm d }\approx-1$), and as a result its perturbations are largely frozen until the onset of the sign-switching epoch, when the background evolution becomes dynamical. Modes that enter the horizon before dark energy becomes dominant begin to decay slowly after horizon entry. In contrast, the smallest-$k$ modes—those that enter the horizon around the sign-switching epoch—grow exponentially after horizon entry. Regarding the velocity of the DE perturbations shown in the bottom-right image of figure~\ref{fig:newtonia_gauge_plots6}, the velocity seems to increasing in the same way for all $k$ modes till matter domination. As different modes enter the horizon they star decreasing. Once the DE density becomes positive the velocity terms bounces back and start increasing.

Finally, in the top-right image of figure~\ref{fig:newtonia_gauge_plots6}, we show the evolution of the full perturbative sector. Overall, a general suppression relative to $\Lambda$CDM is observed across all Fourier modes. The most pronounced deviations arise for intermediate and small $k$ modes; i.e. intermediate or large scale, particularly during the epoch in which the DE component becomes dynamically relevant. In this regime, a Gaussian peak appears due to the background DE quantity $\rho_{\rm d} + P_{\rm d} \propto \exp{\left( \eta^2 ( x-x_\dagger)^2 \right)}$. Once the transition is completed and the DE approaches the dS plateau the perturbations follow a $\Lambda$CDM like behaviour. Despite these features, the overall behaviour remains well controlled throughout cosmic evolution.

\section{Observational Data and methodology\label{sec:methodolofy}}

In this section, we describe the methodology adopted to constrain the cosmological models and the datasets used in the analysis. We perform parameter inference using \texttt{Cobaya}'s Monte-Carlo Markov Chain (MCMC) sampler \cite{Torrado:2020dgo,Neal:2005uqf}, interfaced with a modified version of the Boltzmann solver \texttt{CAMB} \cite{Lewis:1999bs,Howlett_2012}. The theoretical predictions are computed using the \texttt{CAMB} configuration described in Appendix~\ref{appendix:synchronous}.\footnote{ECDM is implemented by modifying the general module of \texttt{EarlyQuintessence} rather than \texttt{DarkEnergyFluid}.} We adopt a Big Bang nucleosynthesis (BBN) consistency relation for the primordial helium abundance using the \texttt{PArthENoPE} 8.80 standard prediction \cite{parthenope880}, which provides a physically motivated prior on the helium fraction consistent with standard BBN calculations. The MCMC convergence is assessed using the Gelman--Rubin criterion, requiring $R-1 < 0.02$ \cite{gelman1992inference}, and the resulting chains are analysed with \texttt{GetDist} \cite{Lewis:2019xzd}. Best-fit and maximum of posterior (MAP) parameters are obtained by maximising the likelihood and the posterior using the minimiser implemented in \texttt{Cobaya}. We employ the default BOBYQA algorithm \cite{cartis2018improvingflexibilityrobustnessmodelbased,Cartis_2021,Powell:2009NA06}.

\begin{table}[t]
    \centering
    \renewcommand{\arraystretch}{1.4} 
    \begin{tabular}{cclc}
    \toprule & \textbf{Param.} & \textbf{  Priors} & \textbf{Distribution}\\
    \toprule 
    \multirow{7}{*}{$\Lambda$CDM parameters}
    & $\Omega_{\rm b0}h^2$   & $\mathcal{U}(0.005,0.1)$  & $\mathcal{N}(0.0224,0.0001)$\\
    & $\Omega_{\rm c0}h^2$  & $\mathcal{U}(0.001,0.99)$ & $\mathcal{N}(0.12, 0.001)$  \\ 
    & $H_0$                 & $\mathcal{U}(20,100)$  & $\mathcal{N}(67,  2)$ \\
    & $\ln{(10^{10}A_s)}$   & $\mathcal{U}(1.61,3.91)$  & $\mathcal{N}(3.05,  0.001)$ \\
    & $n_s$                 & $\mathcal{U}(0.8,1.2)$    & $\mathcal{N}(0.965,  0.004)$\\
    & $\tau_{\rm reio}$     & \quad --                         & $\mathcal{N}(0.051,  0.006)$\\
    \midrule
    \multirow{4}{*}{Extended parameters} 
    & $\log_{10}(\log_{10}\left(1+z_{\dagger}\right))$         & $\mathcal{U}(-0.5214,-0.1555)$ & flat\\
    & $\log_{10}\eta$                & $\mathcal{U}(-1,\log_{10}50)$  & flat\\
    \cdashline{2-4}
    & $z_{\dagger}$         & $\mathcal{U}(1.0,4.0)$ & flat\\
    & $\eta$                & $\mathcal{U}(0.1,50)$  & flat\\
    \bottomrule
\end{tabular}
    \caption{\justifying{Priors adopted for the cosmological parameters used in the MCMC analysis. The quoted ranges correspond to the minimum and maximum values allowed for each parameter, as well as the initial distributions to start MCMC, with $\mathcal N({\rm mean},{\rm std})$ denoting the normal distribution of given mean and standard deviation.
    }}
    \label{tab:param}
\end{table}

For the early-time CMB physics we stick to the baseline 6-dimensional parameter space of the $\Lambda$CDM model $\{\Omega_{\rm b0}h^2$,$\Omega_{\rm c0}h^2$,$H_0$, $\ln\left({10^{10}A_{\rm s}}\right)$, $\tau_{\rm reio}$, $n_{\rm s}\}$, describing respectively the physical baryon and cold dark matter densities,
Hubble constant, amplitude of the primordial scalar power spectrum, the optical depth to reionisation, and the spectral index. In addition, the nuisance parameters $\{T_{\rm cal}, E_{\rm cal}\}$ for the SPT $A_{\rm planck}$ for Planck, and $p_{\rm ACT}$ for ACT are included when considering these likelihoods. For the ECDM model, we choose to parametrise Eq.~\eqref{densityecdm} by the logarithm of the argument inside the error function, i.e.,  $\log_{10}(\log_{10}\left(1+z_{\dagger}\right))$ and $\log_{10}\eta$ to arrive at a linearly separable measure of the transition speed $\eta$, transition point $x_\dagger = -\ln(1+z_\dagger)$, and the asymptotic DE density $1/{\rm Erf}(-\eta x_\dagger)$.  The prior range and distribution for the respective parameters can be found in Table \ref{tab:param}.

\subsection{Data\label{subsec:data}}

We constrain the cosmological parameters using a combination of recent measurements of the CMB temperature and polarisation anisotropies, together with BAO and SNe measurement: 

\textbf{Cosmic Microwave Background (CMB):} We include large-scale temperature information using the Planck 2018 low-multipole likelihood, namely the low-$\ell$ TT power spectrum \cite{Planck:2019nip}. At these scales, the data provide strong constraints on the amplitude and shape of primordial fluctuations. The low-$\ell$ CMB spectrum is also sensitive to DE density as well, indirectly through the integrated Sachs-Wolfe (ISW) effect. At higher multipoles, we adopt a combination of Planck and ACT data, consisting of the ACT DR6 CMB-only likelihood for TT, TE, and EE \cite{AtacamaCosmologyTelescope:2025blo}, together with the ACT--Planck cross-calibrated likelihood \cite{ACT:2023dou,ACT:2023kun,Carron:2022eyg}, which combines ACT measurements with large-scale Planck information and includes a relative calibration parameter to account for residual inter-experiment calibration uncertainties. The multipole ranges are chosen to minimise foreground contamination while retaining the constraining power of small-scale measurements. In addition, we include small-scale polarisation information from SPT-3G, using the first-year temperature and polarisation likelihood in its compressed (“lite”) form \cite{SPT-3G:2025bzu,Balkenhol:2024sbv}. These data extend the CMB power spectra to very high multipoles and significantly improve constraints on parameters affecting Silk damping and small-scale anisotropy structure. We further include CMB lensing information from ACT DR6 and the MUSE reconstruction of the lensing potential power spectrum derived from SPT-3G polarisation data \cite{SPT-3G:2024atg}. These measurements probe the integrated matter distribution along the line of sight and provide additional sensitivity to the growth of structure and late-time cosmology. We refer to this combination of datasets simply as \textsc{CMB}. 

\textbf{Baryon Acoustic Oscillations (BAO):} We include BAO distance measurements from the second data release (DR2) of the Dark Energy Spectroscopic Instrument (DESI) survey \cite{DESI:2025zpo,DESI:2025zgx}. This dataset provides measurements of the transverse and radial BAO scales, expressed in terms of $D_M(z)/r_d$ and $D_H(z)/r_d$, over the redshift range $0.4 < z < 4.2$, together with a low-redshift constraint on $D_V(z)/r_d$ in the range $0.1 < z < 0.4$, where $D_H(z) \equiv c/H(z)$ and $D_V(z) \equiv \left[z D_M^2(z) D_H(z)\right]^{1/3}$. These measurements trace the characteristic clustering scale imprinted by acoustic oscillations in the primordial baryon--photon fluid and act as standard rulers across cosmic time. As a result, they provide robust geometric constraints on the expansion history of the Universe and, when combined with CMB data, significantly tighten limits on DE and late-time cosmological parameters. In the following, we refer to this dataset as \textsc{DESI}.

\textbf{Type I Supernovae (SNe)}: We further incorporate luminosity-distance measurements from the Pantheon+SH0ES supernova compilation \cite{Brout:2022vxf}. This dataset combines a large sample of spectroscopically confirmed Type Ia supernovae spanning a broad redshift range and provides one of the most precise probes of the late-time expansion history. The inclusion of these data allows us to directly constrain the distance–redshift relation at low and intermediate redshifts via the calibration of the standardised absolute magnitude of SNe~Ia by SH0ES\cite{Riess:2021jrx}, thereby tightening constraints on DE parameters when combined with BAO and CMB measurements. We refer to it as either PP if without the calibration or PPS with SH0ES calibration.

\subsection{Statistics\label{subsec:statistics}}

To assess and compare the performance of different models and data combinations, we employ a set of statistical diagnostics based on the posterior distributions obtained from our MCMC analysis. Our framework is organised into two complementary aspects: model selection and data consistency assessment.

For model comparison, we consider a range of information criteria, including the Akaike Information Criterion (AIC)~\cite{Akaike:1974vps} and the Bayesian Information Criterion (BIC)~\cite{Schwarz:1978tpv}, alongside more general Bayesian measures such as the Bayesian evidence, the deviance information criterion (DIC)~\cite{spiegelhalter2002dic}, and the widely applicable information criterion (WAIC)~\cite{watanabe2012widelyapplicablebayesianinformation}. To quantify potential tensions between data sets, we further use indicators such as the Bayesian ratio, goodness-of-fit (GoF), and the suspiciousness statistic~\cite{DES:2020hen,Raveri:2019gdp}. A detailed description of these diagnostics is provided in Appendix~\ref{appendix:statistics}.

For the computation of the statistical quantities reported in Tables~\ref{tab:new} and \ref{tab:tensions}, we adopt the following procedure. For the AIC and BIC, we quote only the mean values, since the best-fit parameters obtained with the \texttt{BOBYQA} minimiser correspond to the maximum-likelihood solution. For the Bayesian probes\footnote{Namely the DIC, WAIC, Bayesian evidence, Bayes factor, and suspiciousness.}, we follow a different prescription. As multiple independent runs are available, each chain is assigned equal weight. The quoted central values and uncertainties are therefore computed from the ensemble of chains according to the definitions given below. The relevant statistical quantities are discussed in Appendix~\ref{appendix:statistics}. The mean value of an observable $\mathcal{O}(\Theta)$ is defined as

\begin{equation}
\langle \mathcal{O}(\Theta) \rangle
=
\frac{1}{N_c}
\sum_{i=1}^{N_c}
\langle \mathcal{O}(\Theta) \rangle_i  \,,
\end{equation}
where $\langle \cdots \rangle_i$ denotes an average over the $i$-th chain and $N_c$ is the total number of chains. The corresponding variance is estimated as

\begin{equation}
\begin{aligned}
\mathrm{var}\!\left(\mathcal{O}(\Theta)\right)
&=
\frac{1}{N_c}
\sum_{i=1}^{N_c}
\left\langle
\left(
\mathcal{O}(\Theta)
-
\langle \mathcal{O}(\Theta) \rangle_i
\right)^2
\right\rangle_i
+
\frac{1}{N_c-1}
\sum_{i=1}^{N_c}
\left(
\langle \mathcal{O}(\Theta) \rangle_i
-
\langle \mathcal{O}(\Theta) \rangle
\right)^2  \,,
\end{aligned}
\end{equation}
where the first term accounts for the variance within each chain, while the second quantifies the variance between chains. For the GoF, the maximum-posterior estimate obtained from the minimiser does not by itself provide an uncertainty. 

In general, these statistics follow a common interpretative structure in which larger values correspond to increasingly disfavoured models or greater tension between data sets, in accordance with Jeffreys’ scale (Table~\ref{table:jeffrey}), with the exception of the GoF and suspiciousness measures.

\section{Observational results\label{sec:observations}}

We have divided the datasets into five combinations:

\begin{itemize}
    \item \textbf{Combination I:} A purely early-universe combination using all the CMB data, Planck, ACT and SPT. We refer to it as just CMB combination.
    \item \textbf{Combination II:} a purely late-universe combination using the PantheonPlus and DESI DR2 data sets (PP+DESI).
    \item \textbf{Combination III:} a purely late-universe combination using the PantheonPlus $\&$ SHOES and DESI DR2 data sets (PPS+DESI).
    \item \textbf{Combination IV:} combined early- and late-time measurements without the SHOES calibration (CMB+PP+DESI).
    \item \textbf{Combination V:} combined early- and late-time measurements with the SHOES calibration (CMB+PPS+DESI).
\end{itemize}

\subsection{Cosmological parameters\label{subsec:parameters}}

\begin{table*}[htbp]
\centering
\scriptsize
\setlength{\tabcolsep}{2.0pt}
\renewcommand{\arraystretch}{1.3}

\begin{tabularx}{\textwidth}{>{\raggedright\arraybackslash}p{2.55cm} *{5}{>{\centering\arraybackslash}X}}
\toprule
& CMB & PP+DESI & PPS+DESI & CMB+PP+DESI & CMB+PPS+DESI \\
\midrule

\multirow{2}{*}{\textbf{Model}}
& ECDM & ECDM & ECDM & ECDM & ECDM \\
& \textcolor{blue}{$\Lambda$CDM} & \textcolor{blue}{$\Lambda$CDM} & \textcolor{blue}{$\Lambda$CDM} & \textcolor{blue}{$\Lambda$CDM} & \textcolor{blue}{$\Lambda$CDM} \\

\midrule

\multirow{2}{*}{$10^2\Omega_\mathrm{b0} h^2$}
& $2.2467\pm0.0098$ & $3.76^{+1.80}_{-0.87}$ & $2.57^{+0.20}_{-0.24}$ & $2.243\pm0.010$ & $2.2464\pm0.0097$ \\
& \textcolor{blue}{$2.241^{+0.000100}_{-0.000090}$}
& \textcolor{blue}{$4.16^{+1.90}_{-0.73}$}
& \textcolor{blue}{$2.86\pm0.15$}
& \textcolor{blue}{$2.2478\pm0.0096$}
& \textcolor{blue}{$2.2532\pm0.0099$} \\

\multirow{2}{*}{$\Omega_\mathrm{c0} h^2$}
& $0.1189\pm0.0011$ & $0.200^{+0.067}_{-0.043}$ & $0.153^{+0.013}_{-0.010}$ & $0.11966^{+0.00068}_{-0.00080}$ & $0.11950\pm0.00086$ \\
& \textcolor{blue}{$0.12036^{+0.00085}_{-0.00120}$}
& \textcolor{blue}{$0.177^{+0.061}_{-0.031}$}
& \textcolor{blue}{$0.1363\pm0.0057$}
& \textcolor{blue}{$0.11819\pm0.00065$}
& \textcolor{blue}{$0.11752\pm0.00065$} \\

\multirow{2}{*}{$100\theta_{\star}$}
& $1.04100\pm0.00026$ & $1.079^{+0.0270}_{-0.0071}$ & $1.0669^{+0.0093}_{-0.0069}$ & $1.04095\pm0.00043$ & $1.04098\pm0.00030$ \\
& \textcolor{blue}{$1.04091\pm0.00026$}
& \textcolor{blue}{$1.068^{+0.0230}_{-0.0042}$}
& \textcolor{blue}{$1.0579^{+0.0054}_{-0.0049}$}
& \textcolor{blue}{$1.04111\pm0.00024$}
& \textcolor{blue}{$1.04118\pm0.00025$} \\

\multirow{2}{*}{$\ln(10^{10}A_\mathrm{s})$}
& $3.032\pm0.012$ & --- & --- & $3.040^{+0.0110}_{-0.0098}$ & $3.041\pm0.011$ \\
& \textcolor{blue}{$3.046^{+0.0120}_{-0.0095}$}
& \textcolor{blue}{---}
& \textcolor{blue}{---}
& \textcolor{blue}{$3.0573\pm0.0099$}
& \textcolor{blue}{$3.060^{+0.0120}_{-0.0080}$} \\

\multirow{2}{*}{$n_\mathrm{s}$}
& $0.9720\pm0.0035$ & --- & --- & $0.9700\pm0.0035$ & $0.9703\pm0.0032$ \\
& \textcolor{blue}{$0.9683\pm0.0040$}
& \textcolor{blue}{---}
& \textcolor{blue}{---}
& \textcolor{blue}{$0.9728\pm0.0030$}
& \textcolor{blue}{$0.9741\pm0.0034$} \\

\multirow{2}{*}{$\tau_\mathrm{reio}$}
& $0.0509\pm0.0058$ & --- & --- & $0.0523\pm0.0069$ & $0.0531\pm0.0058$ \\
& \textcolor{blue}{$0.0542\pm0.0059$}
& \textcolor{blue}{---}
& \textcolor{blue}{---}
& \textcolor{blue}{$0.0584\pm0.0054$}
& \textcolor{blue}{$0.0585^{+0.0074}_{-0.0039}$} \\

\midrule

\multirow{2}{*}{$H_0$}
& $79.2^{+4.4}_{-10.0}$ & $>62.0$ (95\% CL) & $73.6\pm1.0$ & $68.77\pm0.35$ & $69.27^{+0.29}_{-0.39}$ \\
& \textcolor{blue}{$67.25^{+0.49}_{-0.35}$}
& \textcolor{blue}{$>62.0$ (95\% CL)}
& \textcolor{blue}{$73.8\pm1.1$}
& \textcolor{blue}{$68.13\pm0.27$}
& \textcolor{blue}{$68.44\pm0.28$} \\

\multirow{2}{*}{$\Omega_\mathrm{m}$}
& $0.232^{+0.038}_{-0.051}$ & $0.332^{+0.020}_{-0.013}$ & $0.331^{+0.020}_{-0.015}$ & $0.3019\pm0.0038$ & $0.2972\pm0.0036$ \\
& \textcolor{blue}{$0.3172^{+0.0049}_{-0.0073}$}
& \textcolor{blue}{$0.3044\pm0.0079$}
& \textcolor{blue}{$0.3041\pm0.0079$}
& \textcolor{blue}{$0.3044\pm0.0036$}
& \textcolor{blue}{$0.3004\pm0.0037$} \\

\multirow{2}{*}{$\sigma_8$}
& $0.915^{+0.039}_{-0.081}$ & $1.04^{+0.23}_{-0.50}$ & $0.91^{+0.18}_{-0.43}$ & $0.8285\pm0.0060$ & $0.8324^{+0.0056}_{-0.0067}$ \\
& \textcolor{blue}{$0.8156\pm0.0048$}
& \textcolor{blue}{$0.90^{+0.19}_{-0.42}$}
& \textcolor{blue}{$0.80^{+0.16}_{-0.37}$}
& \textcolor{blue}{$0.8144\pm0.0040$}
& \textcolor{blue}{$0.8132^{+0.0046}_{-0.0034}$} \\

\multirow{2}{*}{$S_8$}
& $0.796^{+0.028}_{-0.022}$ & $1.09^{+0.24}_{-0.54}$ & $0.96^{+0.20}_{-0.45}$ & $0.8310\pm0.0077$ & $0.8285\pm0.0078$ \\
& \textcolor{blue}{$0.8387^{+0.0083}_{-0.011}$}
& \textcolor{blue}{$0.90^{+0.19}_{-0.43}$}
& \textcolor{blue}{$0.81^{+0.16}_{-0.38}$}
& \textcolor{blue}{$0.8203\pm0.0068$}
& \textcolor{blue}{$0.8137^{+0.0073}_{-0.0060}$} \\

\midrule

$\log_{10}\log_{10}(1+z_\dagger)$
& $>-0.44\,(95\%~\mathrm{CL})$ & $-0.336^{+0.082}_{-0.11}$ & $-0.328^{+0.091}_{-0.11}$ & $-0.218^{+0.030}_{-0.034}$ & $-0.229^{+0.022}_{-0.040}$ \\

$\log_{10}\eta$
& $<1.16\,(95\%~\mathrm{CL})$ & $0.51^{+0.11}_{-0.19}$ & $0.52^{+0.11}_{-0.21}$ & $0.78^{+0.42}_{-0.36}$ & $0.67^{+0.56}_{-0.50}$ \\

\midrule

$z_{\dagger}$
& $>1.31\,(95\%~\mathrm{CL})$ & $2.02^{+0.39}_{-0.77}$ & $2.08^{+0.45}_{-0.83}$ & $2.93^{+0.21}_{-0.50}$ & $3.07^{+0.30}_{-0.50}$ \\

$\eta$
& $<14.45\,(95\%~\mathrm{CL})$ & $3.76^{+0.10}_{-1.90}$ & $3.903^{+0.034}_{-2.100}$ & $6.5^{+1.9}_{-5.3}$ & $7.6^{+2.2}_{-6.0}$ \\

\midrule

$\Delta$AIC
& $2.602$ & $-3.066$ & $-2.872$ & $-13.058$ & $-3.955$ \\

$\Delta$BIC
& $11.363$ & $7.828$ & $-8.023$ & $-1.5720$ & $7.530$ \\

$\Delta$WAIC
& $-2.224\pm0.701$ & $-0.958\pm0.702$ & $-0.518\pm0.692$ & $-7.982\pm0.612$ & $-15.238\pm0.500$ \\

$\Delta$DIC
& $-1.752\pm0.40$ & $-1.186\pm0.635$ & $-0.395\pm0.549$ & $-7.758\pm0.494$ & $-15.389\pm0.321$ \\

$\Delta(-\ln B)$
& $-2.618\pm0.298$ & $-3.067\pm0.368$ & $-2.695\pm0.389$ & $-8.390\pm0.333$ & $-15.929\pm0.238$ \\

\bottomrule
\end{tabularx}

\caption{\justifying
Mode and $68\%$ confidence interval of the cosmological parameters, late-time observables, and statistical probes obtained for $\Lambda$CDM and ECDM under different dataset combinations. Each parameter is split into two rows, with ECDM shown above in black and $\Lambda$CDM below in blue.
}
\label{tab:new}
\end{table*}

We use \texttt{GetDist} to compute the posterior distributions and contour plots for all cosmological and model parameters. The confidence interval of the cosmological parameters are summarised in Table~\ref{tab:new}, which provides a global comparison between $\Lambda$CDM and the ECDM extension across different dataset combinations. One universal pattern across all dataset combination is the upper bound on the transition speed $\eta$. The data-independent nature indicates the breakdown of the dark energy fluid model that warrants further investigation. For now, let us focus on the data-dependent constraint.

\begin{figure}[t]
    \centering
    \includegraphics[width=0.46\textwidth]{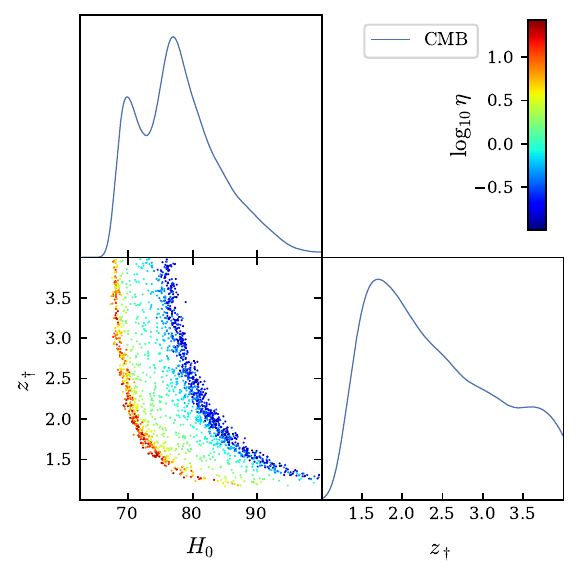}
    \caption{\justifying {68$\%$ and 95$\%$ C.L. posterior distribution of the CMB data for ECDM. We can observe two branches in $H_0$ depending if the transition speed from negative to positive DE density values if fast ($\log_{10}\eta>0$) or slow ($\log_{10}\eta<0$).
    }}
    \label{fig:cmb_branches}
\end{figure}

The CMB dataset alone provides the weakest constraints on the ECDM model parameters, as expected for a scenario in which the modification predominantly affects the late-time expansion history. This is most clearly reflected in the broad posterior distributions of both standard cosmological parameters and the additional ECDM parameters.

A particularly striking feature of the CMB-only analysis is the appearance of a bimodal posterior distribution for the Hubble parameter $H_0$, as shown in figure~\ref{fig:cmb_branches}. This structure reveals the existence of two distinct solution branches. These branches are associated with different values of the transition parameter $\log_{10}\eta$, which controls the rate at which the DE density crosses zero. We identify a slow-transition branch, characterised by $\log_{10}\eta<0$, and a fast-transition branch with $\log_{10}\eta>0$. Both branches provide comparably good fits to the CMB data despite predicting significantly different late-time expansion histories. This indicates that primary CMB observables alone are insufficient to uniquely determine the details of the late-time transition.

To understand the origin of this bimodality, figure~\ref{fig:cmb_branches} shows the posterior distributions in the $(z_\dagger,\log_{10}\eta)$ plane for the CMB dataset. The contours reveal a strong degeneracy between ECDM models when $\eta$ is far enough from unity. On the fast-transition end, the transition becomes nearly instantaneous, and the transition speed barely affects any observable.\footnote{We will analyse the extreme upper bound on $\eta$ later.} On the slow-transition end, DE density $\rho_{\rm d} \approx \rho_{\rm d0}( 1 - \frac{x}{x_\dagger} )$ again is unaffected by the exact transition speed. As a consequence, the CMB likelihood admits an extended region of parameter space rather than a single well-defined solution. The two branches observed in $H_0$ emerge as projections of this higher-dimensional degeneracy onto derived parameters.

\begin{figure}[t]
\centering
\includegraphics[width=\textwidth]{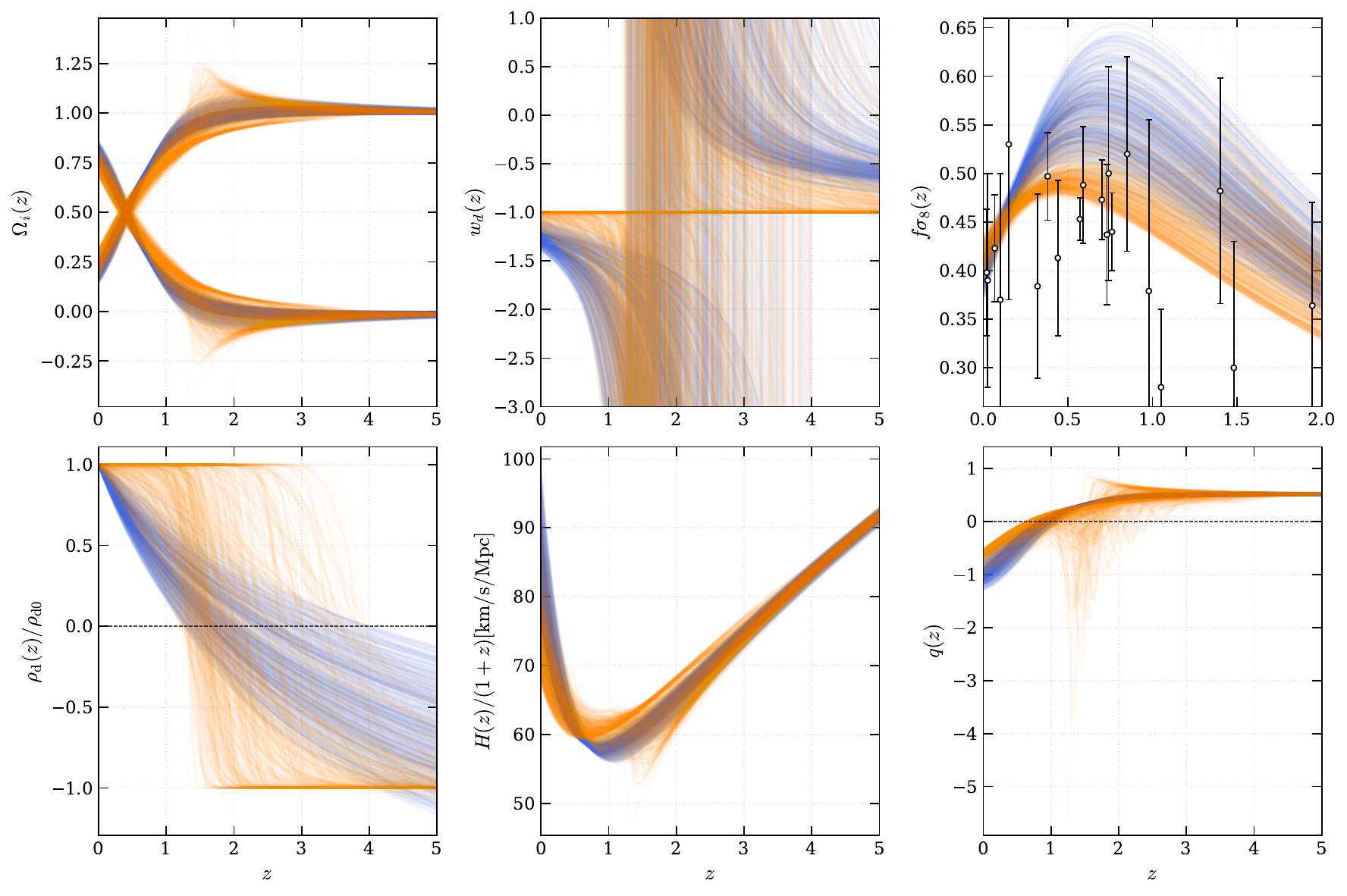}

\caption{\justifying{Evolution of background cosmological quantities as a function of redshift $z$ for the two branches apparent in the CMB combination for the ECDM model. In {\color{BlueViolet}blue} we have the $\log_{10}\eta<0$ branch and in {\color{Orange}orange} the $\log_{10}\eta>0$.
Top-left: fractional energy densities $\Omega_i(z)$ for dark matter and DE. 
Top-centre: DE EoS parameter $w_{\rm d}(z)$. 
Top-right: $f\sigma_8$ where black points with error bars represent observational measurements from redshift-space distortion with data taken from Table~II of \cite{Avila:2022xad}. 
Bottom-left: normalised DE density evolution $\rho_{\rm d}(z)/\rho_{\rm d0}$. 
Bottom-centre: scaled Hubble expansion rate $H(z)/(1+z)$. 
Bottom-right: deceleration parameter $q(z)$, where $q<0$ indicates accelerated expansion.  
Each thin curve corresponds to a random realisation drawn from the posterior distributions, illustrating theoretical uncertainty.
}}
\label{fig:background_branches}
\end{figure}

The existence of multiple viable solutions can be further understood in terms of the background evolution of the Universe. These reconstructions, shown in figure~\ref{fig:background_branches}, illustrate how the two branches correspond to qualitatively different late-time cosmological histories. The slow-transition branch (blue) exhibits DDE behaviour at present time, thus accommodating higher values of $H_0$ and a more extended phase of late-time acceleration, as shown in figure~\ref{fig:background_branches}.The continuously changing DE density at the current epoch also rules it out once SNe data is included.

In the perturbative sector, the growth observable $f\sigma_8$ is found to be only mildly constrained by the current data sets. As illustrated in figure~\ref{fig:background_branches}, both branches of the ECDM model predict an enhancement of the growth amplitude relative to $\Lambda$CDM, although the range of allowed behaviours remains considerably broader in the slow-transition regime. A clearer distinction between the two solutions emerges when comparing their background and perturbative properties. The fast-transition branch ($\log_{10}\eta>0$, orange) yields a DE evolution that is more closely aligned with the expectations of previous analyses~\cite{Ibarra-Uriondo:2026zbp}. In this case, the inferred value of the Hubble parameter is lower than in the slow-transition solution but remains significantly higher than the value derived for $\Lambda$CDM, with a representative estimate of $H_0 = 74.8^{+2.1}_{-6.9} \,{\rm km/s/Mpc}$. This branch also allows for a more intricate late-time expansion history, including a possible additional phase of accelerated expansion.

\begin{figure}[t]
    \centering
    \includegraphics[width=0.83\textwidth]{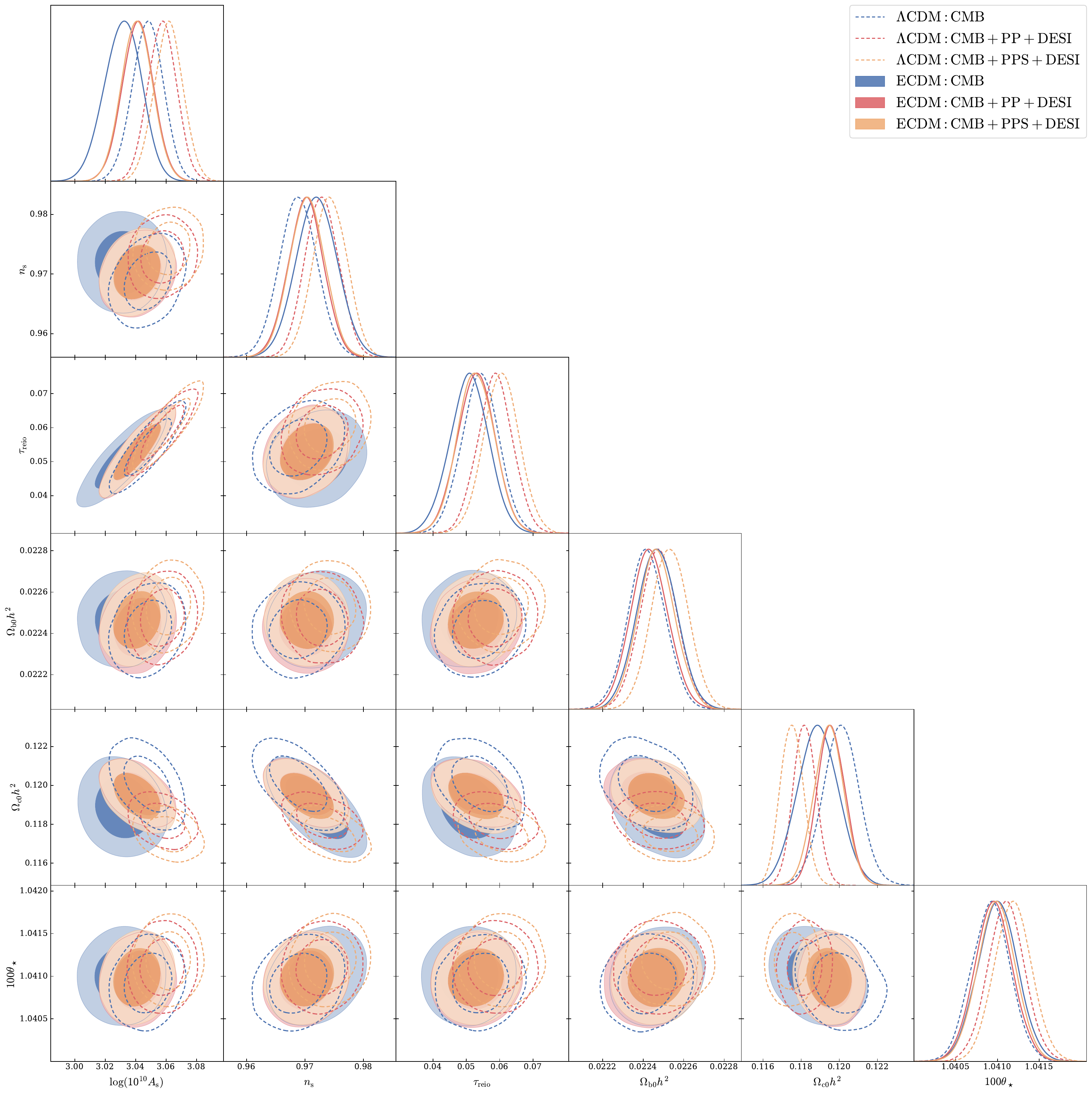}
    \caption{\justifying {68$\%$ and 95$\%$ C.L. posterior distribution of the early-time parameters from
CMB (blue), CMB + DESI + SNe without SHOES calibration (red), CMB + DESI + SNe with SHOES calibration  (orange). Filled contours correspond to ECDM; while dashed ones to $\Lambda$CDM.
    }}
    \label{fig:cmb_quantities}
\end{figure}

By contrast, the slow-transition branch is more strongly affected by parameter degeneracies, which weakens constraints on parameters with $H_0 = 83.2^{+4.1}_{-7.5} \,{\rm km/s/Mpc}$, despite CMB-related observables remaining comparatively stable for both branches under the influence of low-$z$ datasets, as shown in figure~\ref{fig:cmb_quantities}. These degeneracies also impact the inferred growth history: although this branch is in better qualitative agreement with the theoretical expectations of \cite{Bouhmadi-Lopez:2025spo}, it still exhibits a noticeable enhancement of $f\sigma_8$ at low redshifts compared to both $\Lambda$CDM and large-scale structure measurements. All in all, if we just focus on the fast-transition branch, the the Hubble tension is greatly reduced. If we compare $\Lambda$CDM for early vs late Universe dataset we observe that the $5.44\sigma$ tensions decreases down to $0.17\sigma$, though this is accounted due to the high standard deviation of $H_0$ in the ECDM model.\footnote{We quantify the tension for a given parameter $x$ using the standard 1D parameter distance metric, defined as $d = |\bar{x}_{D_1} - \bar{x}_{D_2}| / \sqrt{\sigma_{D_1}^2 + \sigma_{D_2}^2}$, assuming Gaussianity.}

Regarding $S_8$, for CMB data alone, its mean value seems to decrease a lot, thought the standard deviation does increase when compared to $\Lambda$CDM. This is due to the decrease in $\Omega_{\rm m0}$. While $\Omega_{\rm m0}h^2$ remains essentially unchanged with respect to $\Lambda$CDM, the increase in the Hubble parameter leads to a lower inferred matter density, shifting the CMB-only prediction from $S_8=0.839^{+0.008}_{-0.011}$ in $\Lambda$CDM to $S_8=0.796^{+0.028}_{-0.022}$ in ECDM. This reduction moves the early-Universe determination closer to values preferred by weak-lensing surveys and therefore alleviates the overall $S_8$ discrepancy \cite{DES:2026fyc}. Notice that due to different predicted values of $H_0$ by two branches, the derived values of $S_0$ and $\Omega_{\rm m0}$ naturally vary between two branches, with $S_8=0.784_{-0.018}^{+ 0.034}/\Omega_{\rm m0}=0.208^{+0.033}_{-0.028}$ for the slow-transition branch, and $S_8=0.810_{-0.013}^{+0.025}/\Omega_{\rm m0}=0.258^{+0.047}_{-0.022}$ for the fast branch. The tension of $\Omega_{\rm m0}$ between CMB and PPS+DESI thus reduces from $1.8\sigma$ before cutting out the slow branch to $1.43\sigma$ afterwards.\footnote{Given the lack of full-shape information in PP(S)+DESI dataset, the analysis here is only an estimate assuming consistency between BAO observables.} Overall, ECDM tends to lower the CMB-inferred value of $S_8$, bringing it into better agreement with late-time measurements, although this improvement is accompanied by substantially broader parameter uncertainties.

\begin{figure}[t]
    \begin{minipage}{0.44\textwidth}
        \centering
        \includegraphics[width=\linewidth]{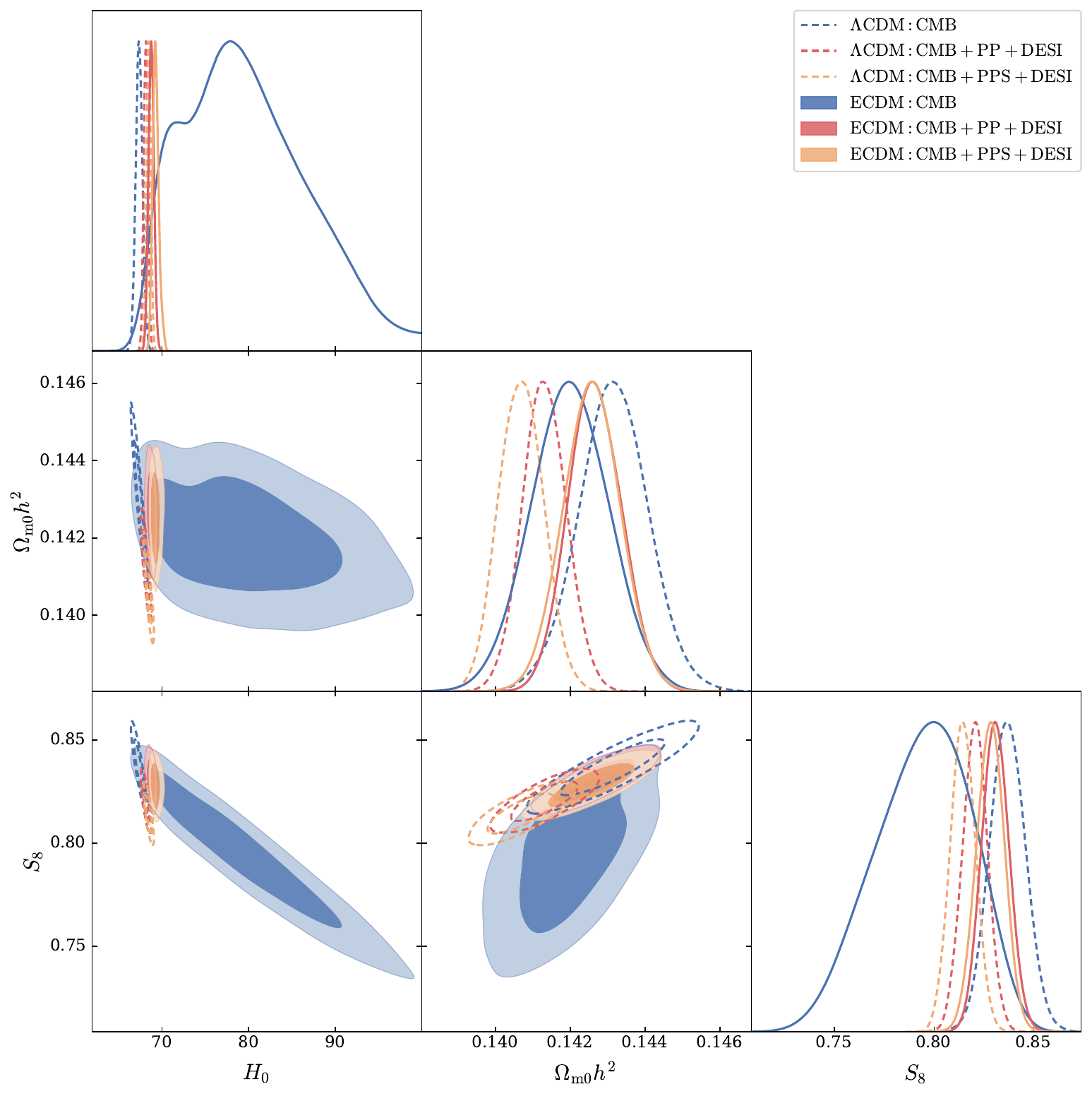}
    \end{minipage}
    \hfill
    \begin{minipage}{0.44\textwidth}
        \centering
        \includegraphics[width=\linewidth]{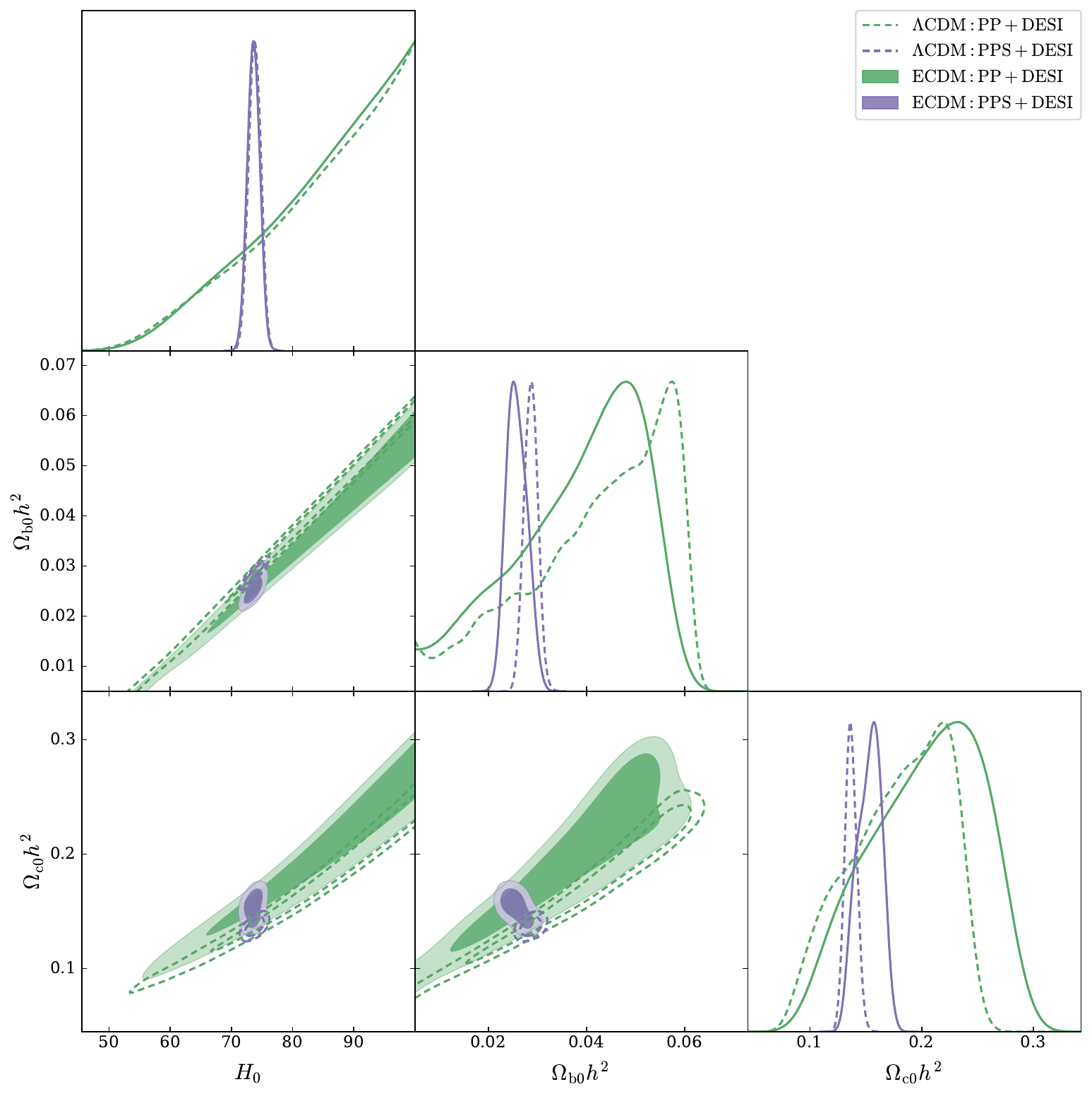}
    \end{minipage}
    \caption{\justifying {
    Marginalised $68\%$ and $95\%$ C.L. two-dimensional posterior distributions for the key cosmological parameters. 
    The left panel displays the parameter space for the Hubble parameter ($H_0$), matter energy density ($\Omega_m$), and clustering amplitude ($S_8$) using early-universe data and full combinations. The right panel shows the constraints on the baryon energy density ($\Omega_b h^2$), cold dark matter energy density ($\Omega_c h^2$), and $H_0$, contrasting the unanchored late-universe probes. In both panels, solid filled contours denote the Extended Dark Energy model (ECDM), while dashed empty contours represent the baseline $\Lambda$CDM framework. The colour coding signifies the dataset combinations: CMB-only (blue), CMB+PP+DESI (red), CMB+PPS+DESI (orange), PP+DESI (green), and PPS+DESI (purple). 
    }}
    \label{fig:tension_contours}
\end{figure}

When utilising BAO and SNe~Ia measurements alone, there is a notable absence of constraints on $\sigma_8$ and $S_8$; this  is expected, given that no a priori constraints were imposed on the primordial  perturbations. For the same reason, $n_{\rm s}$, $\ln(10^{10}A_{\rm s})$, and  $\tau_{\rm reio}$ are omitted from the analysis due to the inherent lack of sensitivity of late-time physics to these parameters. Interestingly, these late-Universe probes alone appear to favour slower and later transitions. This dynamics closely mirrors the behaviour observed in the slow-transition branch derived from the CMB combination, albeit without driving the Hubble constant to such elevated values. The ECDM models seems to lower the baryon density $\Omega_{\rm b0}h^2$ while increasing the amount of cold dark matter $\Omega_{\rm c0}h^2$, increasing the total matter density, as shown in figure~\ref{fig:tension_contours}. 

\begin{figure}[t]
\centering
\includegraphics[width=\textwidth]{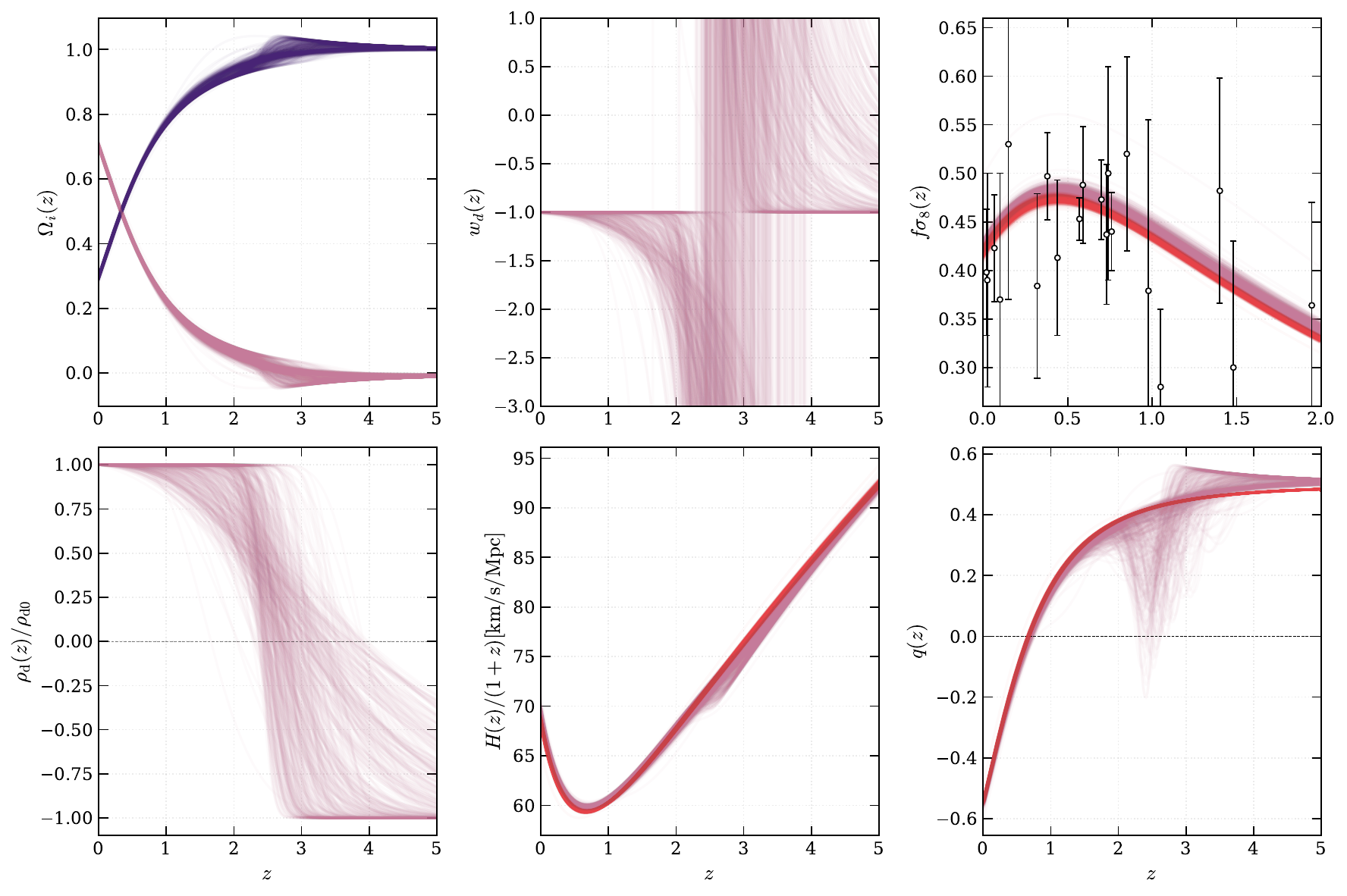}

\caption{\justifying{Evolution of background cosmological quantities for the ECDM model as a function of redshift $z$  for the CMB+PPS+DESI combination, ECDM in {\color{myPink}pink} a $\Lambda$CDM in {\color{myRed}red}.   
Top-left: fractional energy densities $\Omega_i(z)$ for dark matter ({\color{myPurple}purple}) and DE ({\color{myPink}pink}). 
Top-centre: $f\sigma_8$ where black points with error bars represent observational measurements from redshift-space distortion with data taken from Table~II of \cite{Avila:2022xad}. 
Top-right: total effective EoS $w_{\rm tot}(z)$. 
Bottom-left: normalised DE density evolution $\rho_{\rm d}(z)/\rho_{\rm d0}$. 
Bottom-centre: scaled Hubble expansion rate $H(z)/(1+z)$. 
Bottom-right: deceleration parameter $q(z)$, where $q<0$ indicates accelerated expansion.  
Each thin curve corresponds to a random realisation drawn from the posterior distributions, illustrating theoretical uncertainty. 
}}
\label{fig:background_combined}
\end{figure}

Moving on to the early- and late-time combinations, the inclusion of low-redshift BAO measurements from DESI and the standardised candles of PP(S) with CMB data significantly improves the constraints on both standard and ECDM parameters.  These data directly probe the expansion history in the redshift range where DE plays a significant role and therefore strongly reduce the degeneracy present in the CMB-only analysis. As shown in Table~\ref{tab:new}, the uncertainties on $H_0$, $\Omega_m$, and $\sigma_8$ are substantially reduced once DESI + PP(S) data are included. In addition, the posterior distributions become more Gaussian, indicating a significant reduction in model degeneracies. This is specially true for $H_0$, due to tighter constrains on $\log_{10}\eta$, only fast-transition branch remains, and as the result the inferred Hubble constant converges to values similar to what has been reported for the abrupt $\Lambda_{\rm s}$CDM model \cite{Yadav:2025vpx}. In particular, the Hubble parameter shifts towards intermediate values between those preferred by CMB-only ECDM and $\Lambda$CDM. For the full combined datasets, we obtain $H_0 \simeq 69\textrm{ km/s/Mpc}$, which lies closer to local measurements of SH0ES while remaining consistent with early-universe constraints. Similarly, the matter density parameter converges to values around $\Omega_m \simeq 0.30$, in good agreement with large-scale structure observations. The amplitude of matter fluctuations, quantified by $\sigma_8$ and $S_8$, remains broadly consistent with $\Lambda$CDM, with only mild shifts toward slightly higher values in the ECDM framework. Therefore, while the model can alleviate the $H_0$ discrepancy to some extent, it does not simultaneously reduce the predicted clustering amplitude and in fact tends to favour slightly larger values of $S_8$ than the standard model.

\begin{figure}[t]
    \centering

    \begin{minipage}{0.325\textwidth}
        \centering
        \includegraphics[width=\linewidth]{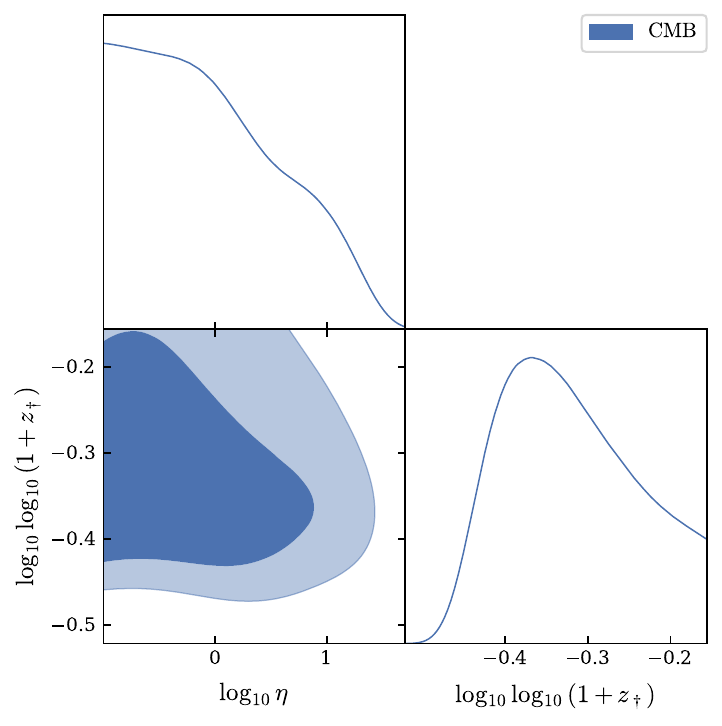}
    \end{minipage}
    \hfill
    \begin{minipage}{0.325\textwidth}
        \centering
        \includegraphics[width=\linewidth]{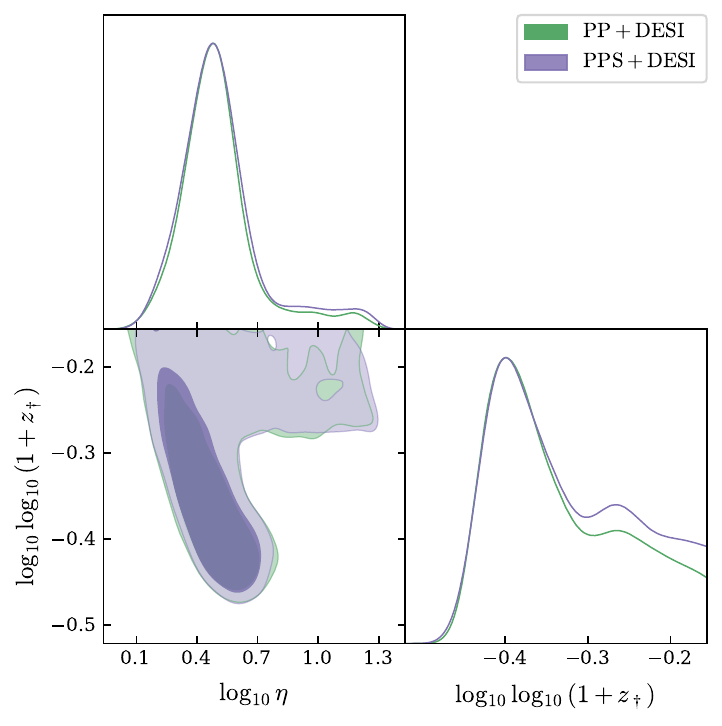}
    \end{minipage}
    \hfill
    \begin{minipage}{0.325\textwidth}
        \centering
        \includegraphics[width=\linewidth]{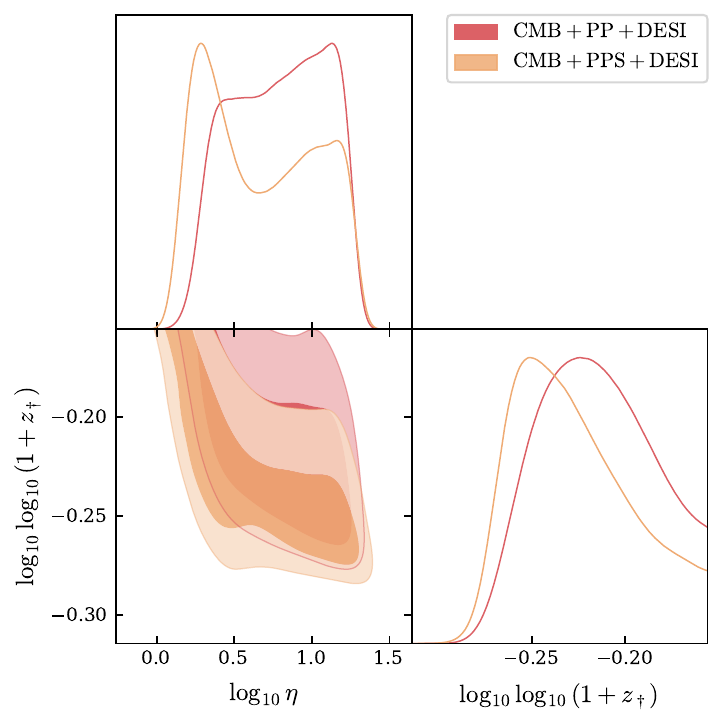}
    \end{minipage}

    \caption{\justifying{68$\%$ and 95$\%$ C.L. posterior distribution for the extra parameters for the ECDM model. On the left figure we observe the early-time measurement alone; the CMB combination in blue. The figure in the middle showcases late-time only measurements; the PP+DESI combination in green and the PPS+DESI combination in purple. The right figure showcases the full combinations; CMB+PP+DESI in red and CMB+PPS+DESI in orange.
    }}

    \label{fig:extra_parameters}
\end{figure}

The improved constraining power is clearly visible in figure~\ref{fig:background_combined}. Whereas figure~\ref{fig:background_branches} exhibits two distinct classes of transition, the addition of the complementary datasets significantly tightens the constraints, eliminating the slow-transition branch and favouring earlier transition with $z_\dagger \gtrsim 2.5$. Similarly, figure~\ref{fig:background_combined} largely excludes the possibility of a third phase of accelerated expansion. In contrast to figure~\ref{fig:background_branches}, where such a feature could not be conclusively ruled out, the preferred reconstructions now favour $q>0$ prior to the onset of the present epoch of accelerated expansion. Furthermore, although the predicted evolution of $f\sigma_8$ converges towards values close to those of $\Lambda$CDM at low redshift, it exhibits a weak but consistent enhancement at earlier times, indicating a stronger growth history,i.e. a slightly faster structure formation than in the standard cosmological model, while remaining consistent with observations.

Figure~\ref{fig:extra_parameters} shows the posterior distributions of the additional ECDM parameters, namely $\left\{ \log_{10} \left( \log_{10}(1+z_\dagger) \right),\, \log_{10}\eta \right\}$.
 The left panel illustrates that CMB data alone are unable to fully constrain this parameter space owing to significant degeneracies with other cosmological parameters. In contrast, the BAO+SNe~Ia combination exhibits a clear preference for a restricted region of the parameter space, providing meaningful constraints on both the transition redshift and its rate. The inclusion of the SH0ES calibration has little impact on the BAO+SNe~Ia constraints, indicating that anchoring the local distance scale does not substantially alter the preferred transition history when only late-time probes are considered, at the cost of shifted baryon and dark matter density, as shown in Figure~\ref{fig:tension_contours}. The situation changes when the full dataset combination is analysed. In this case, the SH0ES calibration shifts the posterior towards larger values of $H_0$, accompanied by a preference for lower values of $z_\dagger$ and a corresponding shift in $\eta$. This behaviour is consistent with the parameter degeneracies present in the model, whereby modifications of the transition history can accommodate larger values of the Hubble constant. Such a trend is qualitatively consistent with the behaviour previously observed in the abrupt $\Lambda_{\rm s}$CDM model, where lower values of $z_\dagger$ increase $H_0$ through their effect on the comoving angular diameter distance. A related mechanism has also been identified in fast-transition sign-switching models, for which increasingly rapid transitions are associated with larger inferred values of $H_0$~\cite{Akarsu:2025gwi}.

\begin{figure}[t]
    \centering 
    \begin{minipage}{0.32\textwidth}
        \centering
        \includegraphics[width=\linewidth]{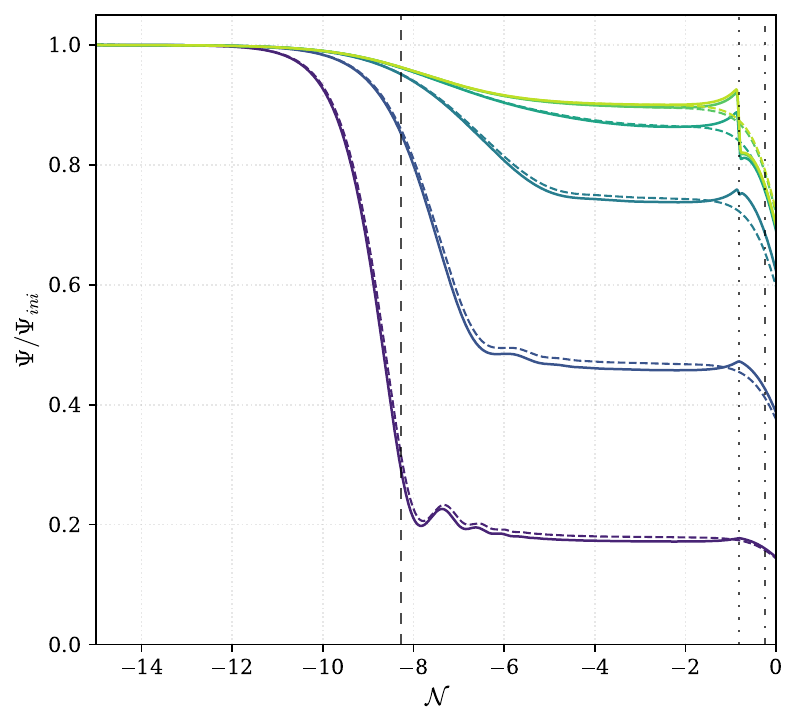}
    \end{minipage}
    \hspace{0.0\textwidth} 
    \begin{minipage}{0.32\textwidth}
        \centering
        \includegraphics[width=\linewidth]{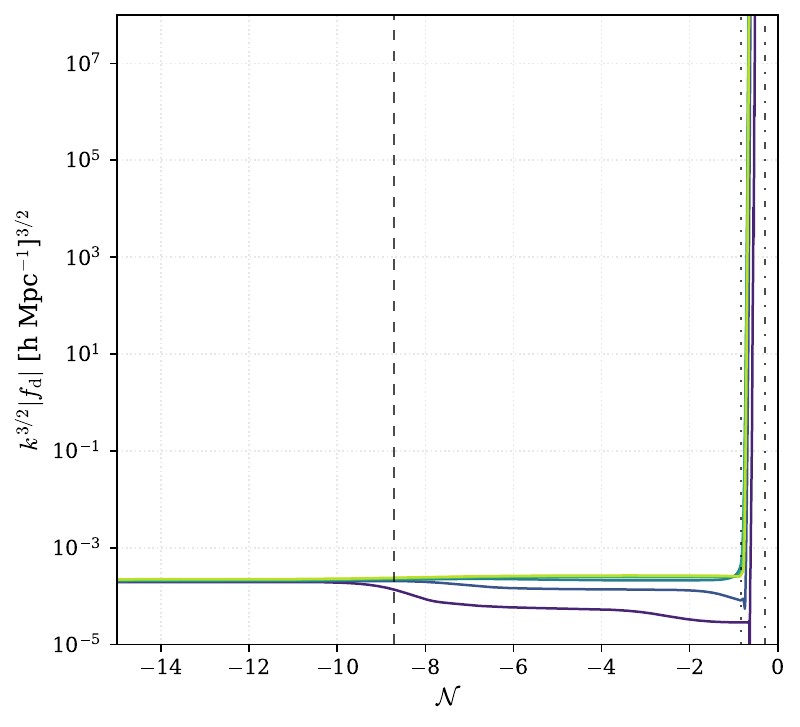}
    \end{minipage}
    \hspace{0.0\textwidth} 
    \begin{minipage}{0.32\textwidth}
        \centering
        \includegraphics[width=\linewidth]{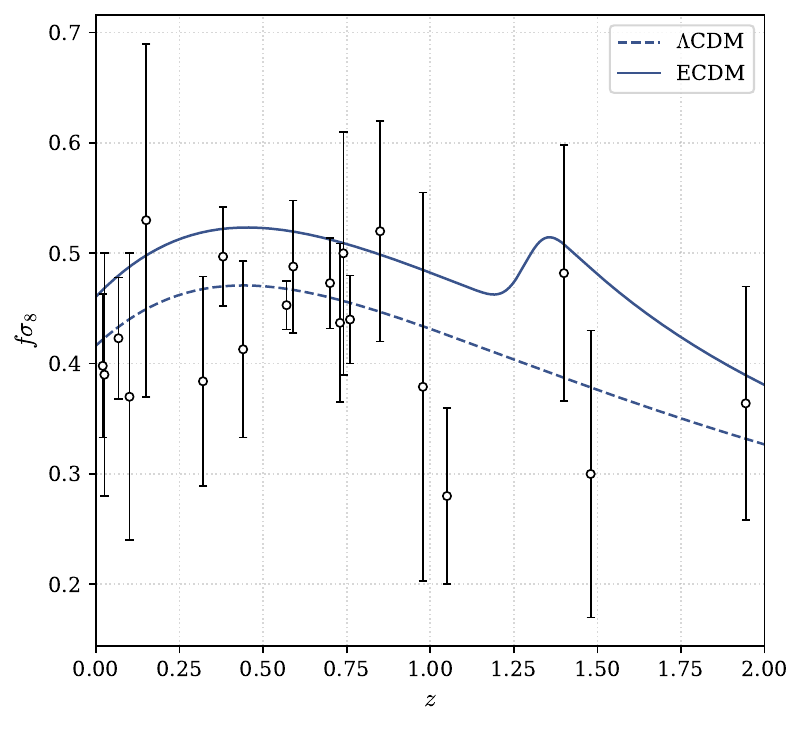}
    \end{minipage}
    
    \caption{\justifying {
Comparison between $\Lambda$CDM (dashed lines) and the ECDM model (solid lines) for four linear perturbation quantities as a function of the e-folding time $\mathcal{N}=\ln a$ for the MAP values form the CMB-SPA+PPS+DESI combination with a fixed transition speed $\eta=10^{3/2}$. The panels show: (left) the gravitational potential $\Psi/\Psi_{\mathrm{ini}}$, (centre) rescaled DE perturbations $k^{3/2}|f_{\rm d}|$, (right) $f\sigma_8$ with error bars represent observational measurements from redshift-space distortion with data taken from Table~II of \cite{Avila:2022xad}.  The Fourier modes are the same as the ones in figure~\ref{fig:newtonia_gauge_plots6}. Vertical black lines indicate characteristic background transitions: radiation–matter equality (dashed), ECDM sign-switching point $x_\dagger$ (dotted), and matter--DE equality (dot-dashed).
}}
    \label{fig:newtonian_gauge3}
\end{figure}

\subsection{Constraint on transition speed\label{subsec:ultra_fast_transition}}

One of the most notable results of this work is the observational disfavouring of extremely rapid transitions. Previous analyses of ECDM~\cite{Ibarra-Uriondo:2026zbp} were unable to establish an upper bound on the transition rate, while related sign-switching models~\cite{Akarsu:2019hmw,Akarsu:2024eoo} have primarily focused on the fast-transition regime. In contrast, our results reveal a markedly different behaviour. For the full CMB+SNe+DESI combination, shown in figure~\ref{fig:background_combined}, the DE density remains dynamically evolving over the redshift interval $2\lesssim z \lesssim 6$, rather than undergoing the abrupt transition characteristic of the $\Lambda_{\rm s}$CDM scenario that sign-switching models are designed to emulate.

To investigate the origin of this preference, we performed an additional analysis in which the transition speed was fixed to the largest value allowed by our prior, namely $\eta=10^{3/2}$. For the CMB+PPS+DESI combination, this choice yields a substantially larger Hubble parameter, $H_0 = 71.16 \pm 0.29\textrm{ km/s/Mpc}$, bringing the model significantly closer to the SH0ES determination. However, this improvement comes at the cost of a pronounced increase in matter clustering, with $S_8 = 0.8948 \pm 0.0054$. This shift is driven by simultaneous increases in both $\Omega_{\rm m0}h^2 = 0.15197 \pm 0.00047$ and $\sigma_8 = 0.8946 \pm 0.0041$. The origin of this behaviour can be traced to the evolution of the perturbations. As the transition becomes increasingly abrupt, both the matter density contrast, $\delta_{\rm m}$, and the gravitational potential, $\Psi$, undergo a rapid variation around the sign-switching epoch observable in figure~\ref{fig:newtonian_gauge3}. This induces a sharp enhancement in the growth rate, reaching values as large as $f\simeq 2.5$, which in turn boosts $f\sigma_8$ well beyond the range favoured by redshift-space distortion measurements. The statistical impact is equally striking: the best-fit $\chi^2$ increases by approximately $10\%$ relative to the unconstrained case. Such a substantial degradation in fit quality provides strong evidence that the current cosmological data do not favour extremely fast transitions in the dark energy sector.

Notice that while it appears that $\log_{10} \eta$ is well constrained in our analysis for every dataset combination, this is simply the bias effect from our choice of prior, as indicated by the unconstrained direction along the upper left corner (orthogonal direction to $\log_{10} \left( -\eta x_\dagger \right)$) in figure~\ref{fig:extra_parameters}. We explicitly verify this claim by extending the upper limit of $z_\dagger$ from $4$ to $10^{2.5}$. The value of $\log_{10} \eta$ shifts accordingly from $0.51^{+0.11}_{-0.19}$ to $0.43^{+0.30}_{-0.22}$ for PP+DESI, cementing this conclusion. On the other hand, the direction along $\log_{10} \left( -\eta x_\dagger \right)$ is highly constrained across all dataset. As mentioned previously, this data-independent constraint originates from the breakdown of the perturbation, which is clearly demonstrated by the oscillatory behaviour of $f_{\rm d}$ after transition.

\subsection{Statistical results\label{subsec:statisticsresults}}

\begin{table}[t]
\centering
\renewcommand{\arraystretch}{1.5} 
\begin{tabular}{lcccc}
\toprule
Test Case & Model & $-\ln{R}$  & $S$  & GoF  \\
\toprule
\multirow{3}{*}{CMB vs. PP+DESI}       & ECDM                           & $5.864\pm0.152$                & $4.357\pm0.093$               & $9.272\pm0.220$  \\
                                       & \textcolor{blue}{$\Lambda$CDM} & {\color{blue} $4.363\pm0.130$} & {\color{blue} $4.439\pm0.038$} & {\color{blue}$3.939\pm0.214$} \\
\cdashline{2-5}
                                       & $\Delta$ & $-1.500\pm0.200$ & $-0.083\pm0.101$ & $5.333\pm0.307$ \\
\midrule
\multirow{3}{*}{CMB vs. PPS+DESI}      & ECDM                           & $25.336\pm0.159$                & $26.790\pm0.082$                & $32.778\pm0.190$  \\
                                       & \textcolor{blue}{$\Lambda$CDM} & {\color{blue} $34.090\pm0.409$} & {\color{blue} $34.143\pm0.137$} & {\color{blue} $33.574\pm0.243$} \\
\cdashline{2-5}
                                       & $\Delta$                       & $-8.753\pm0.439$                & $-7.352\pm0.160$                & $-0.796\pm0.309$ \\
\bottomrule
\end{tabular}
\caption{\justifying{Tension statistics and cross-model comparison for independent dataset blocks. $-\ln R$ is the Evidence Ratio, $S$ is the Suspiciousness, and GoF denotes the Goodness of Fit. The $\Delta$ columns represent the difference between the ECDM and $\Lambda$CDM values for each specific probe ($\Delta = \text{ECDM} - \Lambda\text{CDM}$), quantifying the relative shift in dataset consistency between the two models.}}
\label{tab:tensions}
\end{table}

In this section, we assess how the information criteria and tension probes evaluate the quality of the observational fits and the degree of agreement with the data. The discussion is organised in two parts: first, the information criteria presented in table~\ref{tab:new}, and second, the tension probes.

We further subdivide the discussion of the information criteria into two categories: frequentist and Bayesian approaches. From the frequentist perspective, the AIC appears to favour the ECDM model over $\Lambda$CDM in most cases, with the CMB dataset being the only exception. This behaviour is likely driven by the relatively weaker constraining power of the CMB combination in the context of the ECDM model for this specific analysis. When considering late-time measurements alone, there is a mild weak preference for ECDM, whereas the combination of early- and late-time data leads to a stronger preference, particularly when the SH0ES calibration is excluded. For the BIC, no clear trend emerges. The values fluctuate between moderate preference and moderate disfavour across datasets. Turning to the Bayesian probes, we observe a consistent pattern across the DIC, WAIC, and Bayesian evidence. The CMB combination shows a weak preference for ECDM, while the BAO and SNe~Ia datasets alone provide no decisive evidence in either direction. However, when combining all datasets, the results indicate a strong preference for the ECDM model, especially when the SH0ES calibration is included. Overall, although the frequentist and Bayesian approaches yield slightly different quantitative outcomes, both point towards a general trend favouring the ECDM model over $\Lambda$CDM when data of different epoch are combined.

Table \ref{tab:tensions} compares the tension metrics obtained for ECDM and $\Lambda$CDM using independent dataset combinations. For the CMB versus PP+DESI comparison, ECDM yields a larger Evidence Ratio and Suspiciousness  than $\Lambda$CDM, indicating a stronger disagreement between the datasets. Quantitatively, the differences are $\Delta(-\ln R)=-1.50\pm0.20$ and $\Delta S=-0.08\pm0.10$, where $\Delta=\mathrm{ECDM}-\Lambda\mathrm{CDM}$. The most significant change is observed in the GoF statistic, with $\Delta\mathrm{GoF}=5.33\pm0.31$, suggesting that the ECDM model substantially worsens the joint fit to these datasets relative to $\Lambda$CDM.

In contrast, upon considering the SH0ES calibration, ECDM significantly reduces the inferred tension. Both the Evidence Ratio and Suspiciousness decrease by large amounts, with $\Delta(-\ln R)=-8.75\pm0.44$ and $\Delta S=-7.35\pm0.16$. This indicates that the disagreement present in $\Lambda$CDM is considerably alleviated within the ECDM framework. The corresponding change in the GoF statistic is much smaller, $\Delta\mathrm{GoF}=-0.80\pm0.31$, implying that the tension reduction is not primarily driven by a deterioration of the overall fit quality.

Overall, the results show that ECDM does not uniformly improve dataset consistency. While it substantially reduces the tension between CMB and PPS+DESI, it slightly increases the disagreement between CMB and PP+DESI and leads to a poorer combined fit in that case. This highlights the probe-dependent nature of the model's impact on cosmological dataset concordance, and is consistent with the premise of $\Lambda_s$CDM framework, i.e. to solve the Hubble tension via an abrupt event at late time.

\section{Power spectra\label{sec:powerspectra}}

In cosmology, the statistical properties of the large-scale structure and the CMB are often described in terms of their power spectra. The underlying assumption is that the initial density fluctuations are Gaussian and isotropic, meaning their statistical properties are fully characterised by the two-point correlation function or its Fourier counterpart, the power spectrum.

For a random field $f(\mathbf{x})$, the Fourier transform is defined as
\begin{equation}
\tilde{f}(\mathbf{k}) = \int f(\mathbf{x}) e^{-i\mathbf{k}\cdot\mathbf{x}} \, d^3x  \,,
\end{equation}
and the power spectrum $P_f(k)$ is given by
\begin{equation}
\langle \tilde{f}(\mathbf{k}) \tilde{f}^*(\mathbf{k}') \rangle = (2\pi)^3 \delta_{\rm D}(\mathbf{k} - \mathbf{k}') \, P_f(k)  \,,
\end{equation}
where $\delta_D$ is the Dirac delta function and the angle brackets denote ensemble averages. In practice, $k = |\mathbf{k}|$ is the wavenumber, and $P_f(k)$ quantifies the amplitude of fluctuations at a given scale.

\subsection{Matter power spectrum\label{subsec:matter}}
The matter power spectrum $P_m(k,z)$ describes the distribution of dark matter (and, by extension, total matter) density perturbations $\delta_{\rm m}$
\begin{equation}
\langle \tilde{\delta}_{\rm m}(\mathbf{k},z) \tilde{\delta}_{\rm m}^*(\mathbf{k}',z) \rangle = (2\pi)^3 \delta_D(\mathbf{k} - \mathbf{k}') \, P_{\rm m}(k,z)  \,.
\end{equation}

In linear perturbation theory, the evolution of $\delta_{\rm m}$ decouples into a scale-independent growth factor $D(z)$ and an initial power spectrum from inflation:
\begin{equation}
P_{\rm m}(k,z) = D^2(z) \, P_{\text{init}}(k) \, T^2(k)  \,,
\end{equation}
where $T(k)$ is the transfer function encoding the physics of horizon crossing and matter-radiation equality. On large scales (small $k$), $T(k) \to 1$, and the power spectrum follows the primordial form $P_{\text{init}}(k) \propto k^{n_s}$ with $n_s$ the scalar spectral index. On small scales (large $k$), features such as BAO and the turnover due to the equality scale are imprinted. The non-linear evolution at late times modifies the power spectrum on small scales, which can be described by semi-analytical models like Halofit or emulators.

\begin{figure}[t]
    \centering
    \includegraphics[width=0.6\columnwidth]{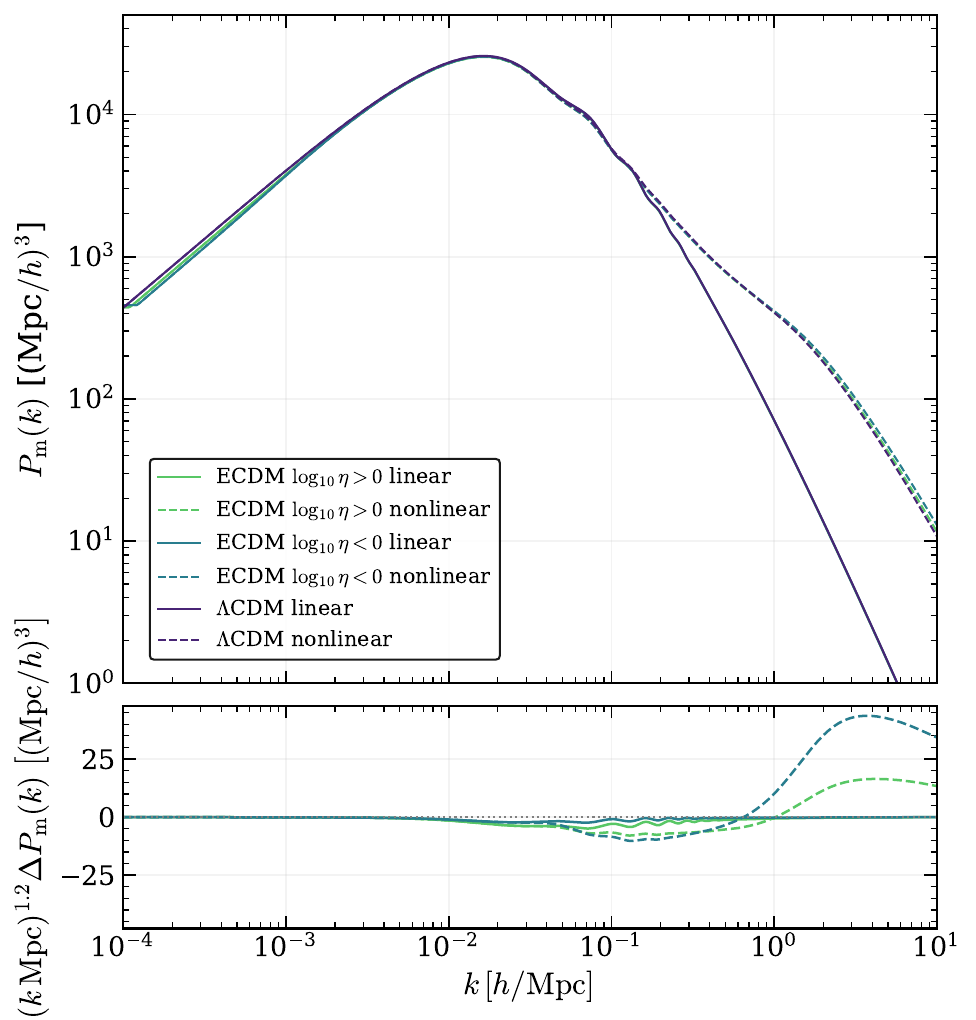}
    \caption{\justifying{Matter power spectrum $P(k)$ for the ECDM and $\Lambda$CDM  cosmologies. The upper panel shows the linear and nonlinear matter power spectra for both models, with solid and dashed lines corresponding to linear and nonlinear predictions, respectively. The lower panel shows the corresponding residuals between ECDM and $\Lambda$CDM, defined as $(k \textrm{ Mpc})^{1.2}\,[P_{\mathrm{ECDM}}(k) - P_{\Lambda\mathrm{CDM}}(k)]$, highlighting the scale-dependent differences in $k$-space.
    }}
    \label{fig:mat_po_spec}
\end{figure}

The behaviour of the present-day matter power spectrum, shown in figure~\ref{fig:mat_po_spec}, exhibits a clear scale-dependent response to the sign-switching DE dynamics. At the intermediate scale, the matter power spectrum is slightly suppressed relative to $\Lambda$CDM over the range $k \sim 10^{-4}$ to $k \sim 8\times 10^{-1}\rm  h/\mathrm{Mpc}$, which should remain compatible with the DESI BAO full-shape analysis\cite{DESI:2024hhd}. Beyond this range, at the large scale the effect of DDE will be discussed later when analysing the CMB lensing, and at the small scale the nonlinear effect enters. While the linear spectrum power of both branches of ECDM and $\Lambda$CDM model converges, the nonlinear spectrum of ECDM transitions to a enhancement regime and diverges from the $\Lambda$CDM prediction, in particular for the slow-transition branch. This small-scale late-time effect may manifest in the weak lensing data of DES \cite{DES:2026zjp}, which will be considered in future study.

The slow-transition branch closely tracks the fast-transition at very large scales and small scales, with a slight deviation around medium scales for the linear evolution. The non-linear case suffer greater deviations for both medium and small scales.

The main differences between the branches therefore originate from the background expansion history and the associated growth rate of matter perturbations. These features reflect the modified late-time growth history induced by the sign-switching DE component, which affects the integrated growth factor while leaving early-time transfer physics largely unchanged.

\subsection{CMB temperature and polarisation spectra\label{subsec:vmb}}

The temperature anisotropies of the CMB are expanded in spherical harmonics:
\begin{equation}
\frac{\Delta T}{T}(\hat{\mathbf{n}}) = \sum_{\ell=1}^{\infty} \sum_{m=-\ell}^{\ell} a_{\ell m} Y_{\ell m}(\hat{\mathbf{n}})  \,,
\end{equation}
where the angular power spectrum is
\begin{equation}
C_\ell^{TT} = \frac{1}{2\ell+1} \sum_{m=-\ell}^{\ell} \langle |a_{\ell m}|^2 \rangle  \,.
\end{equation}
Analogous definitions hold for polarisation ($E$ and $B$ modes) and cross-correlations (e.g., $TE$).

The theoretical $C_\ell$ are computed by solving the Boltzmann equation for photons. They contain a wealth of information: the Sachs–Wolfe plateau at low $\ell$, acoustic oscillations on intermediate scales, and damping at high $\ell$.

\begin{figure*}[t]
    \centering

    \begin{minipage}{0.48\textwidth}
        \centering
        \includegraphics[width=\linewidth]{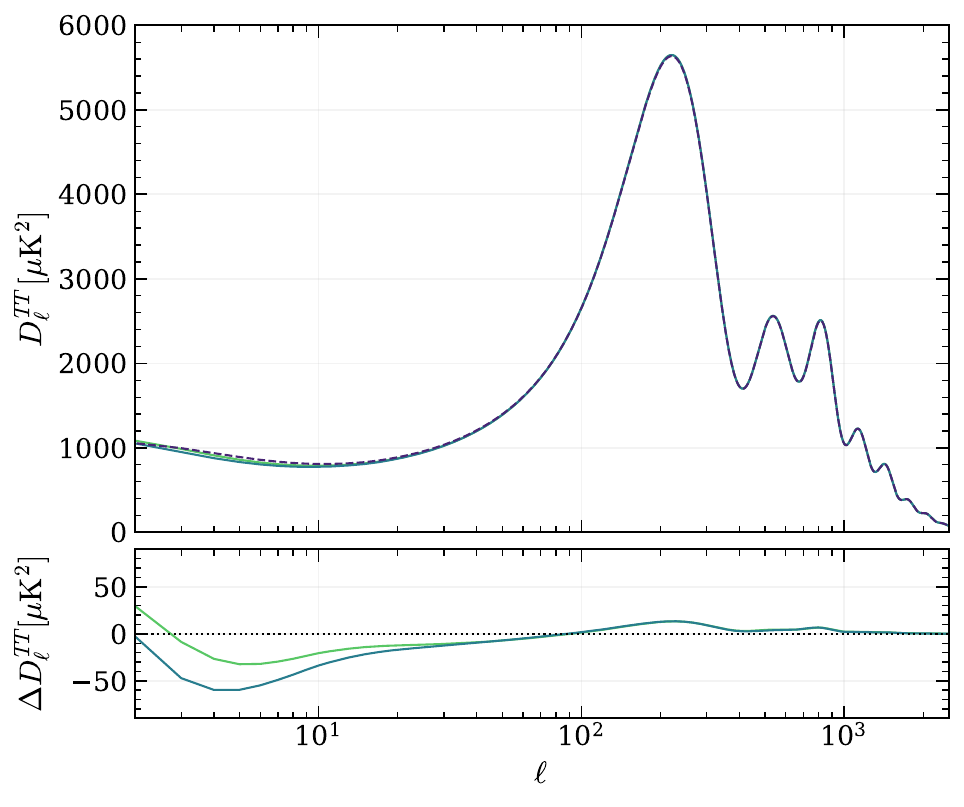}
    \end{minipage}
    \hfill
    \begin{minipage}{0.46\textwidth}
        \centering
        \includegraphics[width=\linewidth]{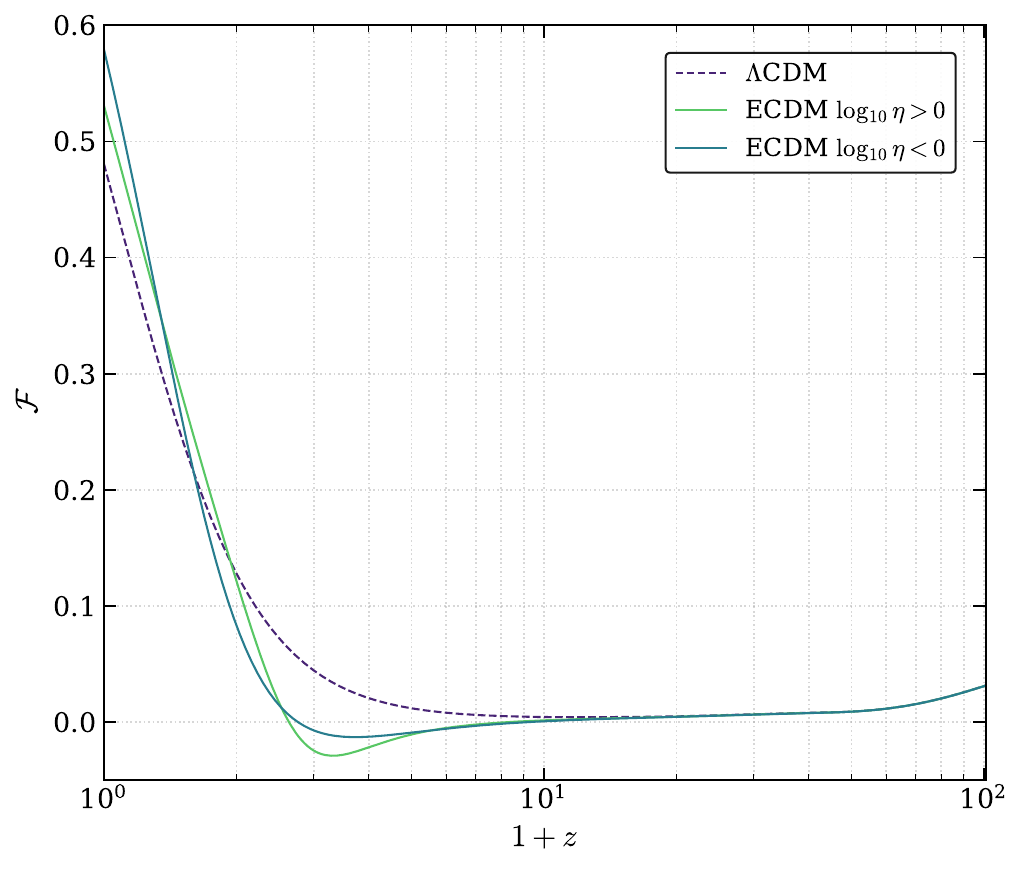}
    \end{minipage}

    \caption{\justifying
    MAP results for (left) the CMB temperature–temperature angular power spectrum ($C_\ell^{TT}$) comparing $\Lambda$CDM (dashed) and ECDM (continuous) models using the CMB-SPA best-fit parameters, and (right) the evolution of $\mathcal{F}(z)$ as a function of $1+z$ used to assess whether the ISW–galaxy cross-correlations are well defined. In the ECDM case, both $\log_{10}\eta>0$ and $\log_{10}\eta<0$ branches are shown and overlap.
    }

    \label{fig:cmb_isw_combined}
\end{figure*}

We now focus on the impact of the model on the CMB temperature anisotropy spectrum $D_\ell^{TT}$. As shown in figure~\ref{fig:cmb_isw_combined}, the characteristic structure of the $\Lambda$CDM spectrum is preserved, including the Sachs--Wolfe plateau at low multipoles, the acoustic peak structure at intermediate scales, and Silk damping at high $\ell$. In both branches of the model, deviations from $\Lambda$CDM are confined predominantly to the largest angular scales, $\ell \lesssim \mathcal{O}(10)$--$\mathcal{O}(30)$. In this regime, a mild suppression of power is observed, reflecting the sensitivity of low multipoles to late-time modifications of the expansion history, in particular through the ISW effect, which will be discussed in detail later. 
At intermediate and high multipoles, the spectrum continues to follow the overall $\Lambda$CDM shape, although small but persistent enhancements remain across a broad range of scales. 
These differences are more moderate than those at low multipoles, but do not fully vanish at higher $\ell$, indicating that the impact of the model extends beyond the largest angular scales while remaining subdominant with respect to the overall acoustic structure. This is reflected by the slight shift in early-time parameters as shown in figure~\ref{fig:cmb_quantities}.

\subsubsection{The Integrated Sachs–Wolfe (ISW) Effect}

The ISW effect \cite{Crittenden:1995ak} originates from the temporal evolution of gravitational potentials linked to large-scale cosmic structures and the expansion dynamics of the Universe \cite{Schaefer:2008qs}. It serves as a direct observational probe of the late-time accelerated expansion and constitutes a secondary anisotropy in the CMB, produced along the photon line of sight.

After recombination, CMB photons propagate freely through the Universe. Deep within the matter-dominated era, the gravitational potential remains nearly constant, as illustrated in figure~\ref{fig:newtonia_gauge_plots6}. Consequently, photons gain energy while falling into a gravitational potential well, but lose the same amount of energy when climbing out of it. Since the potential does not evolve significantly during this epoch, there is no net energy change once the photons leave the well, and therefore no late-time ISW contribution arises. However, as DE begins to dominate and the expansion of the Universe accelerates, the gravitational potentials start to evolve with time. In this case, photons emerging from a potential well may experience a net energy gain or loss, producing additional temperature anisotropies in the CMB through the ISW effect. Although the ISW signal is subdominant relative to the primary CMB anisotropies, it produces a characteristic signature on large angular scales, corresponding to low multipoles, $\ell \lesssim 100$~.

When CMB photons pass through time-varying gravitational potentials along their journey from recombination to the observer, the temperature anisotropies generated due to the energy change from the ISW effect can be expressed as
\begin{equation}
\Theta_{\rm ISW}=\left(\frac{\Delta T}{T_{\rm CMB}}\right)_{\text{ISW}} = \int_{\tau_{\text{rec}}}^{\tau_0} \left( {\Phi'} + {\Psi'} \right) d\tau  \,,
\label{eq:ISW}
\end{equation}
where $T_{CMB}=2.725K$ is the mean CMB temperature. Here we ignore the optical depth of CMB photon for simplicity as it is nearly vanishing in the late universe.

The ISW effect contributes mainly to the lowest multipoles ($\ell \lesssim 10$) of the CMB temperature power spectrum. It also generates a cross-correlation signal between CMB temperature and the large-scale structure traced by galaxies or lensing that constitutes a direct observational probe of DE dynamics and modified gravity scenarios:
\begin{equation}
C_\ell^{T \, \text{g}} \propto \int \frac{dk}{k} \, \Delta^{\text{ISW}}_\ell(k) \, \Delta^{\text{g}}_\ell(k)  \,.
\end{equation}

Although a detailed analysis of the ISW effect lies beyond the scope of this work, we present an approximate estimate of the ISW--galaxy correlation in order to verify that the effect behaves consistently within the model under consideration. Notice that both galaxy counting and CMB lensing observables depend on the Bardeen potentials, it is therefore natural to expect the ``growth factor'' of the potential to appear inside the ratio between $C_\ell^{T{\rm g}}$ and $C_\ell^{\rm gg}$. More precisely, Eq.~\eqref{eq:ISW} can be rewritten as
\begin{equation}
\Theta_{\rm ISW} = 2\int_{\eta_{\text{rec}}}^{\eta_0} aH\left( \Phi + \Psi \right)\mathcal{F} d\eta  \,,
\end{equation}
where $\mathcal{F}=\frac{\partial\ln{(\Phi + \Psi) }}{\partial\mathcal{N}}$ is the potential growth factor \cite{Nakamura:2018oyy}.

In the present model, the DE component is allowed to become negative during part of the cosmic evolution, which in principle could lead to a temporary enhancement of structure formation through its additional attractive contribution to the gravitational potential. As illustrated in figure~\ref{fig:cmb_isw_combined}, $\mathcal{F}<0$ around the sign-switching. This could already be observed in figure~\ref{fig:newtonia_gauge_plots6}, as rather than decaying all the time, the gravitational potential suffers an enhancement around $z_\dagger$, specially for the smallest $k$ modes. This increasing period is directly translated into $\mathcal{F}<0$, so it is expected that a negative ISW-galaxy cross-correlation occurs at a relatively high redshift, followed by a period of enhanced cross-correlation relative to the standard $\Lambda$CDM prediction. This peculiar behaviour aligns well with observational analyses\cite{Stolzner:2017ged,Kovacs:2021mnf}.

\subsection{CMB lensing\label{subsec:lensing}}

Gravitational lensing of the CMB and of galaxies deflects light rays by intervening mass distributions. For the CMB, lensing remaps the primary anisotropies:
\begin{equation}
\tilde{T}(\hat{\mathbf{n}}) = T(\hat{\mathbf{n}} + \nabla \phi(\hat{\mathbf{n}}))  \,,
\end{equation}
where $\phi(\hat{\mathbf{n}})$ is the lensing potential, related to the projected matter density along the line of sight:
\begin{equation}
\phi(\hat{\mathbf{n}}) = -2 \int_0^{\chi_*} d\chi \, \frac{\chi_* - \chi}{\chi_* \chi} \, \Phi(\chi\hat{\mathbf{n}}, \eta_0 - \chi)  \,,
\end{equation}
with $\chi$ the comoving distance and $\chi_*$ the distance to the last scattering surface.

The lensing power spectrum $C_\ell^{\phi\phi}$ is defined analogously to $C_\ell^{TT}$ and is directly related to the integrated matter power spectrum:
\begin{equation}
C_\ell^{\phi\phi} = \frac{9 H_0^4 \Omega_m^2}{4} \int_0^{\chi_*} d\chi \, \frac{(\chi_* - \chi)^2}{\chi_*^2 \chi^2} \, \frac{P_m\left(k = \ell/\chi, z(\chi)\right)}{a^2(\chi)}  \,.
\end{equation}
Observationally, $C_\ell^{\phi\phi}$ can be reconstructed from the non-Gaussianities induced in the CMB fields. It probes the sum of neutrino masses, the amplitude of matter fluctuations ($\sigma_8$), and the late-time expansion history. Additionally, lensing smooths the acoustic peaks in the CMB power spectra and generates a characteristic cross-correlation with the $E$-mode polarisation.

\begin{figure}[t]
    \centering
    \includegraphics[width=0.6\columnwidth]{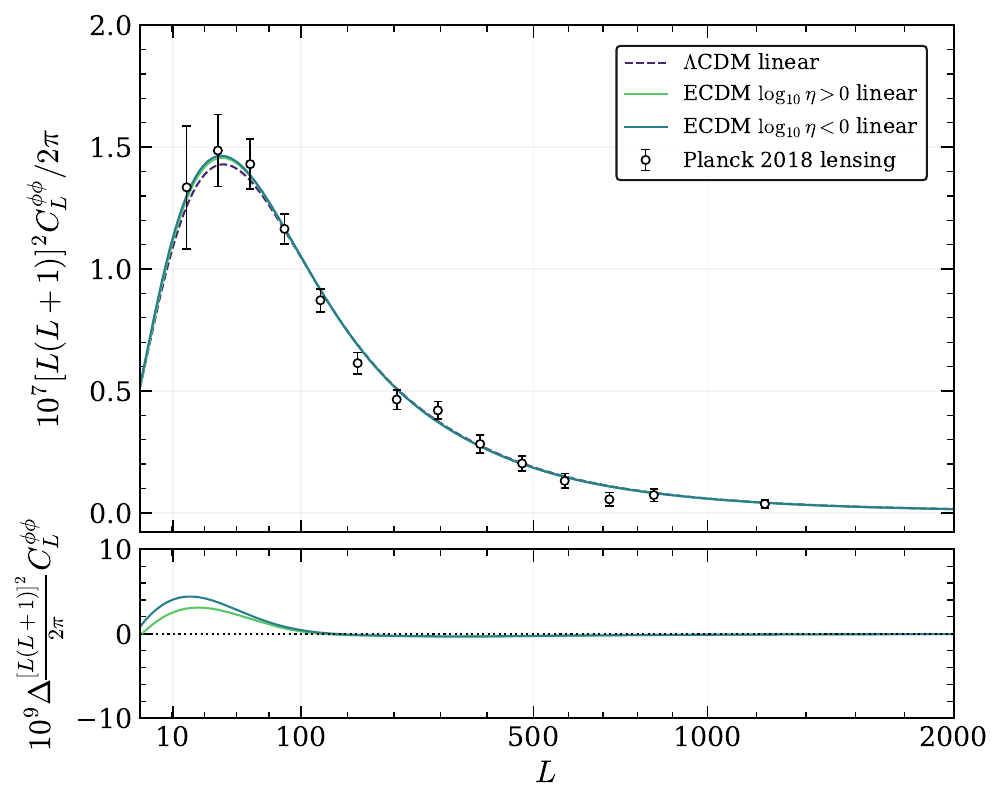}
    \caption{\justifying {MAP value for the Linear Angular power spectrum of the CMB lensing potential $C_L^{\phi\phi}$ (top panel) and absolute residuals (bottom panel) for ECDM model compared to the standard $\Lambda$CDM baseline. Error bars denote the Planck 2018 lensing bandpowers (black data points with $1\sigma$ error bars)~\cite{Planck:2018lbu}. The bottom panel illustrates the residual between the ECDM and $\Lambda$CDM models in units of $10^{9}$.
    }}
    \label{fig:cmb_lensing_comparison}
\end{figure}

We now consider the CMB lensing potential power spectrum $C_{L}^{\phi\phi}$. As shown in figure~\ref{fig:cmb_lensing_comparison}, both branches of the model predict a higher amplitude of the lensing potential relative to the $\Lambda$CDM expectation over a broad range of multipoles. This enhancement slightly favours ECDM over $\Lambda$CDM with $\Delta\chi^2 \sim 0.5$. In both branches, the lensing spectrum exhibits a scale-dependent behaviour: the enhancement becomes more pronounced up to ${L}\sim 25$, after which the spectrum gradually decreases and approaches the $\Lambda$CDM prediction at higher multipoles. This indicates that the largest deviations arise from the largest-scale lensing modes, which are most sensitive to the integrated late-time growth of structure. A slight difference between the two solutions is also observed, with the fast-transition branch predicting a marginally higher amplitude than the slow-transition branch, although the two remain very close overall. Despite these differences, both branches recover a similar trend at high multipoles, where the lensing spectrum becomes increasingly consistent with $\Lambda$CDM.

\section{Conclusions\label{sec:conclusions}}

In this work, we present a comprehensive analysis of the smooth sign-switching ECDM scenario, investigating how current cosmological observations constrain the transition from a negative to a positive DE density when linear DE perturbations are consistently taken into account. This study extends a series of recent investigations of smooth sign-switching DE models and contributes to the broader effort of understanding the observational viability of the sign-switching DE frameworks.

A central aspect of this work is the formulation of a perturbative framework that remains well defined across the sign-switching transition. Conventional treatments of DE perturbations become ill-defined whenever factors proportional to $1+w_{\rm d}$ vanish or diverge, which occurs both in sign-switching scenarios, where $\rho_{\rm d}(z_\dagger)=0$, and in models featuring a crossing of the null-energy-condition boundary, $w_{\rm d}=-1$. To overcome this difficulty, we reformulate the perturbation equations in terms of variables that remain regular throughout the cosmic evolution. The resulting framework provides a consistent description of linear perturbations and can be applied to a broad class of DE models.

In this work we focus on the evolution of the matter density contrast, $\delta_{\rm m}$, the gravitational potential, $\Psi$, the matter growth rate, $f$, and the DE perturbations, $\delta_{\rm d}$, within the ECDM model. We find that the largest-scale modes, namely those entering the horizon around the sign-switching epoch, do not exhibit the suppression in the gravitational potential and growth rate anticipated from previous analyses~\cite{Bouhmadi-Lopez:2025spo}. This difference can be traced back to the inclusion of DE perturbations in the present work. In contrast to Ref.~\cite{Bouhmadi-Lopez:2025spo}, where the DE perturbations are neglected, their inclusion here partially counteracts the expected decay and modifies the evolution of the relevant perturbation quantities around the transition.
In particular, two features stand out as potentially visible within current observational data. First, the matter growth rate is enhanced at the transition across all scales over a period roughly independent of both the transition redshift and the transition speed. This can be directly contested by $f\sigma_8$ data from e.g. redshift-space distortion observations. Second, the gravitational potential at large scale is strongly affected by the DE perturbation at the transition, leading to a modified CMB lensing and ISW spectrum that can be picked up by observation.

To testify the model we utilise a combination of early- and late-time cosmological probes. The data sets are organised into three categories: early-Universe data (CMB alone), late-Universe data (BAO and SNe~Ia, with and without the SH0ES calibration), and their combination. As expected for a late-time modification of the expansion history, the early-Universe data set alone provides relatively weak constraints on the model parameters. Nevertheless, it already exhibits a non-trivial structure in parameter space, favouring two distinct solution branches characterised by different transition behaviours of the DE density. These branches correspond to qualitatively different cosmological histories, typically associated with a slow and a fast sign-switching regime, both of which remain compatible with CMB constraints due to degeneracies with late-time parameters such as $H_0$ and the growth sector, while leaving the early-time physics largely unaffected. When combining the early- and late-time data, the predicted background-related quantities such as the Hubble constant $H_0$ and the transition redshift $z_\dagger$ seem in good agreement with the results of \cite{Ibarra-Uriondo:2026zbp} where the analysis is done at the background level. The transition speed, which ties directly to the perturbation evolution, is now constrained by the CMB spectrum data, with the ultra-fast transition ruled out due to the aforementioned features of the modified perturbation history.

Overall, ECDM provides a better description of the observational data than $\Lambda$CDM according to the relative majority of the statistical criteria considered in this work. Among the information criteria considered, all except the BIC favour ECDM, with the strongest preference arising for the full dataset combination including the SH0ES calibration. Even the early-time CMB only data is neutral between ECDM and $\Lambda$CDM (with the exception of BIC) despite the inclusion of two extended parameters. This is reflected by an enhanced CMB lensing spectrum that matches better with observation. Furthermore, it hints at a distinctive correlation history between CMB ISW and matter density observables, which aligns with observations\cite{Stolzner:2017ged,Kovacs:2021mnf}. The tension metrics lead to a broadly consistent conclusion. Although ECDM does not improve the consistency between the CMB and PP+DESI datasets, the inclusion of the SH0ES calibration strengthens the preference for the sign-switching scenario relative to the cosmological constant, yielding a notable reduction in the tensions associated with the full dataset combinations.

An important conclusion of this work is that arbitrarily fast transitions appear to be strongly disfavoured once the full perturbative sector is taken into account, a regime that has not been explored in previous analyses of sign-switching DE models. Nevertheless, it remains unclear whether this result reflects a generic feature of sign-switching cosmologies or is instead specific to the error-function parametrisation adopted here. Addressing this question requires moving beyond the present phenomenological framework towards a more fundamental description, such as a scalar-field realisation capable of consistently modelling and constraining the transition dynamics. Such an investigation lies beyond the scope of the present work and will be pursued in future studies.

\section*{Acknowledgements}

The authors are grateful to Luis Escamilla for helpful discussions and insights related to this project. M.~B.-L. is supported by the Basque Foundation for Science, Ikerbasque. Our work is supported by the Spanish Grant PID2023-149016NB-I00 funded by MCIN/\allowbreak AEI/\allowbreak 10.13039/\allowbreak 501100011033 and by the ERDF, “A way of making Europe”. It is also supported by Basque Government Grant No.~IT1628-22 (Spain); in particular, B.~I.~U. is funded through this grant. H.-W.~C. is supported by NSFC Grants No.~12250410250, No.~12505074 and No.~12347133. This article is based upon work from COST Action CA21136, “Addressing observational tensions in cosmology with systematics and fundamental physics” (CosmoVerse), supported by COST (European Cooperation in Science and Technology).

\appendix

\section{Cosmological perturbation theory and the gauge transform}\label{appendix:CPT}

Having an inhomogeneous matter distribution, the Einstein equations imply inhomogeneous metric perturbations about the spatially flat FLRW metric. The FLRW line element up to linear perturbations in the scalar sector can be written as \cite{Bassett:2005xm}

\begin{equation}
\begin{aligned}
ds^2 = a^2 \big{[} &- (1+2A) d\tau^2 + 2\partial_i B dx^i d\tau  \\
&+ ( (1-2\psi) \gamma_{ij} + 2\partial_{ij} E ) dx^i dx^j \,\big{]}  \,,
\end{aligned}
\label{eq:perturbed_flrw}
\end{equation}
where $\partial_i \equiv \partial / \partial x^i$ denotes the spatial partial derivative and $\gamma_{ij}$ is the metric on the 3-dimensional space with constant curvature $\mathcal{K}$. The four degrees of freedom in the scalar sector are encoded in the potentials $A = A(\tau,\vec{x})$, $B = B(\tau,\vec{x})$, $\psi = \psi(\tau,\vec{x})$, and $E = E(\tau,\vec{x})$. From Eq.~\eqref{eq:perturbed_flrw}, we can then express the perturbations of the metric and its inverse\footnote{ We use $\delta [g^{-1}]^{\mu\nu}$ to refer to the linear perturbation of the inverse metric $g^{\mu\nu}$, which at first order in perturbation differs from the perturbation of the metric raised indices by a difference in sign: $\delta [g^{-1}]^{\mu\nu}=-\delta g^{\mu\nu}$.}

\begin{equation}\delta g_{\mu\nu} = \begin{pmatrix}
-2A & a\partial_iB \\
a\partial_iB & -2a^2(\psi\gamma_{ij}-\partial_i\partial_jE)
\end{pmatrix},
\end{equation}
\begin{equation}
\delta [g^{-1}]^{\mu\nu} = \begin{pmatrix}
2A^{-2} & a^{-1}\partial^iB \\
a^{-1}\partial^{i}B & 2a^{-2}(\psi\gamma^{ij}-\partial^i\partial^jE)
\end{pmatrix},
\end{equation}
where the Latin indices are raised by the inverse of the $3$-dimensional metric.

The unit 4-vector normal to the spatial hypersurface $n^{\mu}$ reads \cite{Malik:2001rm}
\begin{equation}
n_{\mu}=-(1+A,\textbf{0})  \,,\quad
n^\mu=\left(1-A,-\frac{\partial^iB}{a}\right)  \,.
\label{eq:unit 4-vec}
\end{equation}

From Eq.~\eqref{eq:unit 4-vec} we can compute the perturbed expansion scalar $\theta$, the shear tensor $\sigma_{ij}$ and the acceleration vector $a_{\mu}$ in a perturbed FLRW universe as \cite{Malik:2001rm}
\begin{equation}
    \begin{aligned}
        \theta & \coloneqq 3H-\frac{3}{a}\left(\psi'+\mathcal{H}A\right)+\frac{1}{a^2}\nabla^2\sigma  \,,\\
        \sigma_{ij} & \coloneqq\left(\partial_i\partial_j-\frac{\gamma_{ij}}{3}\nabla^2\right)\sigma  \,,\\
        a_i&\coloneqq\partial_iA  \,,
    \end{aligned}
    \label{eq:quantities}
\end{equation}
where $\nabla^2\coloneqq\gamma^{ij}\partial_{ij}$ is the spatial Laplacian operator. In Eq. (\ref{eq:quantities}) we have introduced the scalar shear potential
\begin{equation}
    \sigma\coloneqq a\left( E'-B\right)  \,.
    \label{eq:shear}
\end{equation}

The intrinsic Ricci scalar curvature of constant time hypersurface is given by 
\begin{equation}
    ^{(3)}R=\frac{4}{a^2}\nabla^2\psi  \,,
\end{equation}
and its perturbation
\begin{equation}
    \delta^{(3)}R=\frac{4}{a^2}(\nabla^2+3\mathcal{K})\psi  \,.
\end{equation}
Due to these expressions we refer to $\psi$ as the curvature perturbation.

Under scalar coordinate/gauge transformation
\begin{equation}
    \begin{aligned}
        \tau&\rightarrow \tau+\delta \tau  \,,\\
        x^{i}&\rightarrow x^{i}+\gamma^{ij}\partial_{j}\delta x  \,,
    \end{aligned}
\end{equation}
$ \delta \tau$ and $\delta x$ determine the time slicing and spatial threading respectively. The four scalar potentials encode the perturbations of the metric and change as follows under gauge transformations
\begin{equation}
    \begin{aligned}
        A&\rightarrow A-\delta \tau'  \,,\\
        B&\rightarrow B+\delta \tau+a{\delta x}'  \,,\\
        E&\rightarrow E-\delta x  \,,\\
        \psi & \rightarrow\psi+\mathcal{H}\delta \tau  \,.
    \end{aligned}
\end{equation}
Notice that $A$ and $\psi$ are gauge independent with respect to spatial gauge transformations. In the case of $B$ and $E$, although they are spatially gauge dependent, the same is not true the for the shear defined in Eq. (\ref{eq:shear}), which is spatially gauge independent and described the shear potential for the anisotropic shear wordlines orthogonal to constant time surfaces \cite{Kodama:1984ziu}.
There are only two gauge-invariant potentials that can be constructed using only geometrical quantities
\begin{equation}
    \begin{aligned}
        \Phi&\coloneqq A-\frac{\sigma'}{a}  \,,\\
        \Psi&\coloneqq \psi+\mathcal{H}\frac{\sigma}{a}  \,.
    \end{aligned}
\end{equation}
These are known as the Bardeen potentials \cite{Bardeen:1980kt}, and they reduce to the scalar potentials $A$ and $\psi$ in the orthogonal zero-shear/Newtonian/longitudinal gauge, defined by $E = B = 0$ (which also implies $\sigma = 0$). Many readers may be more familiar with the Newtonian gauge for the theoretical description of perturbations; accordingly, we adopt this framework in Section~\ref{sec:metric} to analyse the stability of the perturbations from a theoretical perspective.

Nevertheless, for observational applications, the Einstein--Boltzmann codes most widely used in the literature, \texttt{CAMB} and \texttt{CLASS}, are formulated in the synchronous gauge. For this reason, readers may also find the same approach in the corresponding gauge on Appendix~\ref{appendix:synchronous}.

\section{Synchronous gauge\label{appendix:synchronous}}

The synchronous gauge is defined by setting $A = B = 0$, with\cite{Ma:1995ey}
\begin{equation}
    ds^2 = a^2 \left[ d\tau^2 - \left( ( 1 - 2\psi ) \gamma_{ij} + 2 \partial_i \partial_j E \,\right) dx^i dx^j \right]  \,,
\end{equation}
and for those more used to the \texttt{CAMB} notation we introduce the synchronous gauge variables $h = -2k^2 E - 6\psi$ and $\eta = \psi$ where $h$ is the trace of 3-dimensional metric perturbation 
$h_{ij}=-2\psi\gamma_{ij}+2\partial_i\partial_jE$
\begin{equation}
    ds^2=a^2\left[-d\tau^2+(\gamma_{ij}+h_{ij})dx^idx^j\right]  \,,
\end{equation}
where $h_{ij}$ represents the spatial metric perturbations. We will be working in the Fourier space $k$ in this paper. We introduce two fields $h(\mathbf{k}, \tau)$ and $\eta(\mathbf{k}, \tau)$ in $k$-space and write the scalar mode of $h_{i j}$ as a Fourier integral

\begin{equation}
    \begin{aligned}
    h_{i j}(\vec{x}, \tau)=&\int d^3 k e^{i \mathbf{k} \cdot \vec{x}}\left\{\hat{k}_i \hat{k}_j h(\mathbf{k}, \tau) \right. \\
    &\left.
    +\left(\hat{k}_i \hat{k}_j-\frac{1}{3} \delta_{i j}\right) 6 \eta(\mathbf{k}, \tau)\right\}, \quad \mathbf{k}=k \hat{k}  \,.
    \end{aligned}
\end{equation}

Note that $h$ is used to denote the trace of $h_{i j}$ in both the real space and the Fourier space. Same as before, following a perfect fluid approach, we can compute the perturbations for the different components\footnote{Notice that while the $(- , + , + , +)$ metric signature and the definition of ${T_A}^{i}_{\,0} = ( \rho_A + P_A ) \partial^i v_A$ is adopted for the Newtonian gauge calculations, the synchronous gauge analysis employs the $(+ , - , - , -)$ signature and the definition of $\hat k^i {T_A}^0_{\,i} = ( \rho_A + P_A ) v_A$ to maintain consistency with the underlying framework of \texttt{CAMB}. Two sign flips cancel out, and we can simply rescale $v$ by $k$ when translating between two systems.}

\begin{equation}
    \begin{aligned}
    \delta_A'&=-3 \mathcal{H}\left({c}_{s A}^2-w_A\right)\left(\delta_A+3 \mathcal{H}\left(1+w_A\right) v_A / k\right)\\
    &-\left(1+w_A\right) k v_A-3 \mathcal{H} w_A' v_A / k-3\left(1+w_A\right) h'  \,,\\ 
    v_A'&=-\mathcal{H}\left(1-3 {c}_{s A}^2\right) v_A-k \sigma-k {c}_{sA}^2 \delta_A /\left(1+w_A\right)  \,.
    \end{aligned}
    \label{syngauge}
\end{equation}

Under this notations we can define the shear as $\sigma=(h-3\eta)/2k$. Following the same procedure as in the Newtonian gauge we rewrite Eq.~\eqref{syngauge} in terms of Eq.~\eqref{eq:new_pert}

\be\begin{aligned}
f_A' =& -3 \mathcal{H} \left( c_{s A}^2 - c_{aA}^2 \right) \left( f_A + 3 \mathcal{H} v_A / k \right)  - kv_A + 3 h'  \,,\\
v_A' =& - \mathcal{H} \left( 1 - 3 c_{s A}^2 \right) v_A - k \sigma + k c_{s A}^2 f_A  \,.
\end{aligned}\ee

To relate the metric perturbation in two gauges, we perform an infinitesimal coordinate shift from the synchronous coordinates ($x^\mu$) to the Newtonian coordinates ($\hat{x}^\mu$):
\begin{equation}
    \hat{\eta} = \eta + \alpha(\eta, \mathbf{x})  \,,\quad
    \hat{x}^i = x^i + \nabla^i \beta(\eta, \mathbf{x})  \,.
\end{equation}

By transforming the metric tensor component-by-component ($\hat{g}_{\mu\nu} = g_{\alpha\beta} \frac{\partial x^\alpha}{\partial \hat{x}^\mu} \frac{\partial x^\beta}{\partial \hat{x}^\nu}$) and working in Fourier space ($\nabla^2 \to -k^2$), matching the components yields these fundamental equations for the coordinate functions $\alpha$ and $\beta$:
\begin{equation}
    \beta' - \alpha = 0 \implies \alpha = \beta'  \,,\quad
    h' + 6\eta' = 2k^2 \alpha  \,.
\end{equation}

For the Newtonian Potential $\Phi$ (from $g_{00}$):

\begin{equation}
    \Phi = \alpha' + \mathcal{H}\alpha  \,.
\end{equation}

Substituting our expression for $\alpha$ into this equation, we apply the product rule to get the relation using first and second conformal time derivatives:
\begin{equation}
    \Phi= \frac{1}{2k^2} \left[ h^{^{{\prime\prime}}} + 6\eta^{\prime\prime} + \mathcal{H}(h' + 6\eta') \right]  .
\end{equation}

For the Spatial Curvature Potential $\Psi$ (from $g_{ij}$):
\begin{equation}
    \Psi = \eta - \mathcal{H}\alpha  \,.
\end{equation}

Substituting $\alpha$ gives the spatial relationship:
\begin{equation}
    \Psi = \eta - \frac{\mathcal{H}}{2k^2}(h' + 6\eta')  \,.
\end{equation}

\section{Statistical probes\label{appendix:statistics}}

In this work, the statistical indicators employed can be broadly divided into two categories. The first consists of model-selection and model-comparison metrics, which quantify the balance between goodness of fit and model complexity. These are given by

\begin{table}
\centering
\renewcommand{\arraystretch}{1.4} 
\begin{tabular}{ccl}
\toprule
$\Delta(-\ln{B})$ & \textbf{Interpretation} \\
\midrule
$> 10$ & Strongly disfavored / tensioned \\
$5 \sim 10$ & Moderately disfavored / tensioned \\
$2 \sim 5$ & Weakly disfavored / tensioned \\
$-2 \sim 2$ & Inconlusive \\
$-5 \sim-2$ & Weakly favored / aligned \\
$-10 \sim-5$ & Moderately favored / aligned \\
$< -10$ & Strongly favored / aligned \\
\bottomrule
\end{tabular}
\caption{\justifying{Jeffreys' scale for deciding the evidence of model $M_1$ over $M_2$ or the tension between datasets.}}
\label{table:jeffrey}
\end{table}

\begin{equation}
    \begin{aligned}
\mathrm{AIC}(D \mid M) & \equiv -2\ln P\left(D \mid \theta_{M,\mathrm{bf}}\right)+2u  \:,\\
\mathrm{BIC}(D \mid M) & \equiv -2\ln P\left(D \mid \theta_{M,\mathrm{bf}}\right)+k\ln N  \,,\\
-\ln B(D \mid M) & \equiv -2\ln \int P\left(D \mid \theta_M\right) P\left(\theta_M\right) d \theta_M \\
&=-2\ln V_M-2\ln \left\langle\left(P\left(D \mid \theta_M\right)\right)^{-1}\right\rangle_{M \mid D}  \;\,,\\
\mathrm{DIC}(D \mid M) & \equiv 2\ln P\left(D \mid \theta_{M, \text { map }}\right)-4 F(D \mid M)  \:,\\
\mathrm{WAIC}(D \mid M) & \equiv-2F(D \mid M)+\mathrm{BMD}(D \mid M)  \:,\\
F(D \mid M) & \equiv\left\langle\ln P\left(D \mid \theta_M\right)\right\rangle_{M \mid D} \equiv \ln B(D \mid M)+\mathrm{KL}(D \mid M)  \:,\\
\mathrm{BMD}(D \mid M) & \equiv 2\left(\left\langle\left(\ln P\left(D \mid \theta_M\right)\right)^2\right\rangle_{M \mid D}-\left\langle\ln P\left(D \mid \theta_M\right)\right\rangle_{M \mid D}^2\right)  ,\\
    \end{aligned}
\end{equation}

In the expressions above, $D$ denotes the observational dataset and $M$ the cosmological model under consideration. The set of model parameters is represented by $\theta_M$. The quantities $P(D|\theta_M)$, $P(\theta_M)$, and $P(\theta_M|D)={P(D|\theta_M)P(\theta_M)}/{B(D|M)},$
correspond to the likelihood, prior, and posterior distributions, respectively. Posterior averages are defined as
\[
\langle \cdots \rangle_{M|D}
=
\int (\cdots)\,P(\theta_M|D)\,d\theta_M,
\]
while
\[
V_M=\int P(\theta_M)\,d\theta_M,
\]
denotes the prior volume. The quantity $F$ is the Bayesian deviance, $\mathrm{KL}(D|M)$ is the Kullback--Leibler divergence between posterior and prior distributions, and $\mathrm{BMD}$ denotes the Bayesian model dimension. The subscript ``map'' refers to the maximum-a-posteriori estimate of the parameters and ``bf'' top the best-fit value. Finally, $u$ represents the number of free parameters in the model and $N$ the effective number of data points entering the likelihood.\footnote{ECDM carries 2 extra parameters, i.e., $\Delta u = 2$. For BIC, $N = 1701$ for PP(S), $N = 14$ for DESI and $N=590$ for CMB.} To interpret these probes, we follow Jeffreys’s scale \cite{Jeffreys:1939xee}, summarised in Tab. \ref{table:jeffrey}.

The AIC and BIC provide simple penalised-likelihood estimators that reward a good fit while discouraging unnecessary model complexity. In both cases, smaller values indicate a statistically preferred model. Model comparisons are therefore performed through differences with respect to a chosen reference scenario, typically $\Lambda$CDM. 

The Bayesian evidence can be estimated directly from the posterior samples using the harmonic-mean identity given above. However, it is well known that this estimator may become numerically unstable due to the excessive weight assigned to samples located in the low-likelihood tails of the posterior distribution. To reduce this sensitivity, we consider a stabilised truncated harmonic-mean estimator in which the evidence is evaluated only within a restricted high-posterior region. The corresponding expression is

\begin{equation}
    \begin{aligned}
        \ln B(D \mid M)_{\alpha} & \equiv 2\ln \alpha -2 \ln V_M  - 2\ln \left\langle\left(P\left(D \mid \theta_M\right)\right)^{-1}\right\rangle_{M \mid D, \theta \in \Theta_{\alpha}}  \;, \\
        \alpha & \equiv \int_{\Theta_{\alpha}} P\left(\theta_M \mid D\right) d \theta_M  \,,
    \end{aligned}
\end{equation}

where $\Theta_{\alpha}$ denotes the subset of parameter space satisfying $ P(D|\theta_M)\geq \mathcal{L}_{\rm min}$
with $\mathcal{L}_{\rm min}$ chosen such that the enclosed posterior mass equals $\alpha$. The expectation value is then computed using only samples belonging to $\Theta_{\alpha}$. Throughout this work we adopt $\alpha=0.95$, retaining $95\%$ of the posterior volume while excluding the extreme tails responsible for the instability of the standard harmonic-mean estimator. The additive correction term $\ln\alpha$ accounts for the discarded posterior fraction. 

The second category of statistical indicators consists of tension estimators, designed to quantify the level of agreement between independent datasets. In a simpler way, it will let us know the inherent tension between datasets $D_1$ and $D_2$ for model model $M$. These quantities are defined as

\begin{equation}
    \begin{aligned}
-\ln R\left(D_1, D_2 \mid M\right)  \equiv&-\ln B\left(D_1 D_2 \mid M\right)+\ln B\left(D_1 \mid M\right)+\ln B\left(D_2 \mid M\right), \\
\mathrm{GoF}\left(D_1, D_2 \mid M\right)  \equiv&-2\ln P\left(D_1 D_2 \mid \theta_{M, \text { map }}\right) \\
& +2\ln P\left(D_1 \mid \theta_{M, \text { map }}\right)+2\ln P\left(D_2 \mid \theta_{M, \text { map }}\right), \\
S\left(D_1, D_2 \mid M\right)  \equiv&-2F\left(D_1 D_2 \mid M\right)+2F\left(D_1 \mid M\right)+2F\left(D_2 \mid M\right).
\end{aligned}
\end{equation}

Here, $R$ is the Bayesian evidence ratio, $\mathrm{GoF}$ measures the relative goodness of fit between the combined and individual datasets, and $S$ is the suspiciousness statistic, which isolates disagreement between datasets after removing the influence of prior volume effects.

\bibliographystyle{JHEP}

\bibliography{bibliography.bib}

\end{document}